\documentstyle[12pt,aaspp4]{article}
\tighten

\begin{document}


\def\zz{\hang\noindent}

\def\kms{km s$^{-1}$}
\def\pix{pix$^{-1}$}
\def\deg{$^\circ$}

\def\mic{{$\mu$m}}

\def\h2o{H$_2$O}

\def\ak{{\it $A_K$}}

\def\teff{$T_{\rm eff}$}

\def\aple{$\mathrel{\hbox{\rlap{\hbox{\lower4pt\hbox{$\sim$}}}\hbox{$<$}}}$}
\def\apge{$\mathrel{\hbox{\rlap{\hbox{\lower4pt\hbox{$\sim$}}}\hbox{$>$}}}$}


\title{Really Cool Stars at the Galactic Center}

\author{R. D. Blum\altaffilmark{1,2}}
\affil{JILA, University of Colorado\\Campus Box 440, Boulder, CO,
80309\\rblum@casa.colorado.edu}

\author{K. Sellgren\altaffilmark{1,3} and D. L. DePoy\altaffilmark{1}}
\affil{Department of Astronomy, The Ohio State University\\
 174 W. 18th Ave., Columbus, Oh, 43210\\
sellgren@payne.mps.ohio-state.edu\\
depoy@payne.mps.ohio-state.edu}

\altaffiltext{1}{Visiting Astronomer, Cerro Tololo Inter--American Observatory,
National Optical Astronomy Observatories, which are operated by the Association
of Universities for Research in Astronomy, Inc., under cooperative
agreement with the National Science Foundation.}

\altaffiltext{2}{Hubble Fellow}

\altaffiltext{3}{Alfred P. Sloan Research Fellow}


\begin{abstract}
New and existing $K-$band spectra for
19 Galactic center late--type stars have been analyzed along
with representative spectra of disk and bulge M giants and supergiants.
Absorption strengths for strong atomic and molecular features have
been measured.
The Galactic center stars generally exhibit stronger absorption
features centered near Na I (2.206 \mic) and Ca I
(2.264 \mic) than representative disk
M stars at the same CO absorption strength.

Based on the absolute $K-$band magnitudes and CO and \h2o \ absorption
strengths for the
Galactic center stars and known M supergiants and asymptotic giant
branch (AGB) stars, we conclude
that only IRS 7 must be a supergiant. Two other bright stars in
our Galactic center sample are likely supergiants as well.
The remaining bright, cool
stars in the Galactic center that we have observed
are most consistent with being
intermediate mass/age AGB stars.
We identify five of the Galactic center stars as long period variables
based on their $K-$band spectral properties and associated
photometric variability. Estimates of initial masses and ages for the
GC stars suggest multiple epochs of star formation have occurred in the
Galactic center over the last 7--100 Myr.

\end{abstract}


\section{INTRODUCTION}
Blum et al. (1996, hereafter Paper I)
presented near infrared photometry for the Galactic center (GC)
stellar population (within the central $\sim$ 4--5 pc).
Analysis of the $JHKL$ photometry showed an excess component
of bright stars in the stellar population compared to the population
in the nearby bulge field
known as Baade's window (BW, $l, b =$ 1\deg,$-$4\deg).

This excess of bright stars has been known for some time (see the
discussion in Paper I and references therein)
and is
generally attributed to recent star formation resulting in
younger, more massive stars than exist in BW.
Substantial evidence for very recent star formation (\aple 8 Myr) has come
from studies of hot emission--line stars in the GC
(Forrest et al. 1987; Allen et al.
1990; Krabbe et al. 1991; Libonate et al. 1995; Blum et al. 1995$a,b$;
Krabbe et al. 1995; Figer 1995; Tamblyn et al.
1996). These stars are thought to be post-main-sequence objects with
initial masses $>$ 35 M$_\odot$. But, as pointed out in Paper I, based
on published spectra and the spectra we present here, the
emission--line stars are not the most conspicuous stars in the GC
bright component. The brightest GC stars at $K$
are largely stars identified as being cool M
stars.

An important distinction between the cool and hot stars is
that the luminous cool stars can trace the most recent epochs of star formation
(red supergiants) as well as older ones (M giants and intermediate age
asymptotic giant branch stars), while the hot stars trace only the former.
In this paper, we extend the work begun by Lebofsky et al. (1982,
hereafter LRT) and Sellgren et al. (1987, hereafter S87) which sought
to identify the most luminous M stars in the GC as massive red
supergiants or less massive giants and so trace star formation there.
Our new spectroscopy (this paper) and photometry (Paper I) are of higher
angular
resolution than these earlier studies. The spectra presented here
sample GC stars in the brightest five magnitudes of the observed
$K-$band luminosity function (Paper I).
We will also discuss the GC
cool stars in the context of asymptotic giant branch (AGB) stars,
the most luminous of which are long period variables (LPVs).
This is particularly relevant in light of the fact that photometric
variables have recently been identified in the GC (Haller 1992;
Tamura et al. 1994, 1996; Paper I).

\section{OBSERVATIONS AND DATA REDUCTION}
The GC observations were obtained on the nights of 1993 July 11$-$13 on the
4--m telescope at the Cerro Tololo Inter--American Observatory (CTIO)
using the Ohio State Infrared Imager and Spectrometer
(OSIRIS). Spectra of three disk M giants were obtained with OSIRIS on the
CTIO 4--m during the night of 1994 June 28. OSIRIS is described
by DePoy et al. (1993).
All basic data reduction procedures were accomplished using
IRAF\setcounter{footnote}{3}.\footnote{IRAF is distributed by the National
Optical Astronomy Observatories.}

Observations and data reduction of the OSIRIS GC $K-$band spectra
taken in 1993 July
have been discussed in detail by Blum et al. (1995b). Briefly, these
$R$ ($=\lambda/\Delta\lambda$) $\sim$ 570 (19.4 \AA \ \pix) spectra were
extracted
from 0.4$''$ pix$^{-1}$ long slit images (slit oriented E--W)
of the central
102$''\times10''$ of the Galaxy
(slit width $\sim$ 1.2''). We also obtained long slit images centered
on several additional stars up to $\sim$ 30$''$ from the center.
The GC spectra were obtained in seeing of $\sim$ 1$''-1.5''$.
Each frame was flat--fielded and sky
subtracted; this included a secondary flat--field to
account for scattered light (see Blum et al. 1995$b$).
The sky images were taken at positions 500$'' - 600''$
off the GC. In addition, local background apertures were also used to
subtract diffuse emission and the underlying stellar background. Each
extracted spectrum was also divided by the spectrum of an A or B star
to correct for telluric absorption. Prior to this correction,
the Br$\gamma$ (2.16 \mic)
absorption in the
A and B stars was removed by estimating a continuum across it by eye.

Spectra for one supergiant (1993 July) and
three disk M giants (1994 June) listed in Table 1
were obtained with the same instrumental setup as the GC
OSIRIS spectra. The supergiant RT Car was
observed with a neutral density filter.
The reduction was similar to the GC stars with
the exception that no secondary flat--field was used
for the 1994 June data
since additional baffling in OSIRIS and improved observing
procedures (see the discussion
on the high resolution spectrum of IRS 13 in Blum et al. 1995$b$)
eliminated the effects of scattered light.

Table 1 also includes
spectra of three disk supergiant stars, five disk M giants, and GC stars
from Kleinmann \& Hall (1986; hereafter KH) and
S87 which have
been re--binned to the same resolution as the OSIRIS data. These high
resolution spectra were first re--binned to the 19.4 \AA \ pix$^{-1}$
sampling of the OSIRIS spectra, then each data point was replaced with
the weighted average of its two neighbors and itself (weights: 0.5, 0.5,
1.0). A spectrum of the GC star IRS 24 (19 \AA \ \pix, not re--binned)
was kindly provided by D. Levine and D. Figer and included in our sample. This
spectrum was previously published by Levine et. al (1995).
The GC star and disk star spectra taken with OSIRIS and the IRS~7,
IRS~23, IRS 24, and $\mu$ Cep
spectra are shown in Figure~\ref{spect}.
All spectra in this figure have been de--reddened according to the $A_K$
values given in Table 4 of Paper I
and are presented as the normalized
ratio of the stellar spectrum to the hot star spectrum used in
correcting telluric absorption.

We also make use of spectra of
14 field M giants and 10 BW M giants from Terndrup et al. (1991),
three supergiants from Hanson et al. (1996),
and FTS spectra
of five Mira stars (LPVs) and one M supergiant from Johnson \& M\'{e}ndez
(1970).
All these stars are listed in Table~1.
These spectra (22 \AA \ pix$^{-1}$, 16 \AA \ pix$^{-1}$,
and $\sim$ 40 \AA \ pix$^{-1}$,
respectively) were not re--binned. The bulge and LPV stars were
normalized in the same way
as the GC and disk stars before analysis. A plot of the
Johnson \& M\'{e}ndez spectrum of R Cas (M7e, Mira)
is shown in Figure~\ref{spect}.

The disk giants and supergiants were used to compare measured atomic
absorption feature strengths to similar measures in the GC
stars. Equivalent widths (W$_{\lambda}$) were measured in bands centered
near the Na I doublet ($\lambda$ $\approx$ 2.206 \mic, $\Delta\lambda$ = 0.015
\mic)
and the Ca I triplet ($\lambda$ $\approx$
2.264 \mic, $\Delta\lambda$ = 0.013 \mic).
A linear continuum was calculated across each feature by interpolating
between nearby continuum positions on either side of each feature.
Inspection of the high resolution and re--binned KH and S87 spectra shows
that the W$_{\lambda}$ for both the Na I and
Ca I features have contributions from other
absorption features.
Hereafter, we will refer to the atomic measurements as ``Na'' and ``Ca''
due to the contributions of multiple species to the absorption strength
(see \S~3.3 for details of the additional contributors).
Due to the more coarse sampling of the Johnson \& M\'{e}ndez (1970)
stars and their generally lower
signal--to--noise, we did not attempt to make ``Na'' and ``Ca'' measurements
for
them. Instead, we supplemented two of these stars with measurements of
``Na'' and ``Ca'' from the high resolution atlas of Wallace \& Hinkle (1996).
The spectra for these two stars ($o$ Cet and $\alpha$ Ori)
were re-binned to the OSIRIS resolution.

In addition to the atomic line measures, we computed absorption strength
measurements for the CO bandhead at 2.2935 \mic \ and for \h2o \
near the blue end of the spectra.
Both of these measures were computed as the percentage of flux in the
band relative to a continuum band at 2.284 \mic \ ([1 $-$ $F_{band}/F_{cont}]
\times 100$). All three fluxes were
measured in 0.015 \mic \ wide bands. The CO band was placed
with its center at 2.302 \mic \ and the \h2o \
band was centered at 2.095 \mic. Our CO and \h2o \ indices are not the same
as the more well known narrow--band photometric indices (e.g., Frogel et al.
1978), but they are correlated with the photometric indices.
See the discussion below on the correlation between the our indices and
those of KH which have, inturn,
been shown by KH to be correlated with the narrow--band indices.

The uncertainties reported in Table 1 were derived by
taking the
rms deviation in regions between features as the uncertainty in a single
pixel and then propagating this uncertainty through the
definitions of the absorption measures.
The absorption strengths were calculated after de--reddening the
spectra. Each GC spectrum was de-reddened using the estimates of $A_K$
in Table 4 of Paper I and the interstellar extinction curve of Mathis
(1990). The BW stars were de-reddened by
$A_K$ = 0.14 mag (Frogel \& Whitford 1987, hereafter FW87); none of the disk
giants or LPVs include correction for
extinction. The reddening of the supergiants
was estimated either from their observed colors (Hoffleit 1982; Nicolet 1978)
and assumed intrinsic colors (Johnson 1966), or was adopted from Elias et al.
(1986, hereafter EFH).  We corrected the supergiants for extinction values
of $A_K$ = 0.1 -- 0.6 using a Mathis (1990) extinction curve with $R$ = 3.1.

The atomic absorption feature equivalent widths
are unaffected by extinction. The CO strengths are affected
only marginally since the bands are close together.
The formal uncertainty in the CO strength due to uncertainty in
$A_K$ can be expressed as $\Delta$CO/$\Delta$$A_K$ $=$ 1.0 $\times$
(1 $-$ CO) $\%$ mag$^{-1}$. The CO strength is underestimated if
$A_K$ is underestimated.
The \h2o \ strengths are more dependent on the derived
reddening because of the large wavelength difference between the flux
and continuum bands. In this case, $\Delta$\h2o/$\Delta$$A_K$ $=$ $-$13.8
$\times$ (1 $-$ \h2o) $\%$ mag$^{-1}$. The \h2o \ strength is overestimated
if $A_K$ is underestimated.

The re--binned disk star spectra show similar absorption strengths as the
disk stars of the same spectral type which have spectra taken on the
OSIRIS system (see Table 1),
suggesting that one--to--one comparisons can be made for ``Na,'' ``Ca,'' and
CO absorption
strengths measured on these different systems.
Comparison of
similar spectral type data data for disk stars from KH  and
this paper to the disk stars from Terndrup et al. (1991), however, shows
slightly redder continua.
This means that the \h2o \ values for the
Terndrup et al. (1991) disk stars may be a few percent larger than the
other disk stars of similar spectral type in our sample.
We do not consider the difference
in observed \h2o \ strengths significant, however, due to the
small number of stars in each sample. For M5 -- M7 giants,
the Terndrup et al. giants have a mean \h2o \ of 10.4 $\%$ $\pm$
0.9 $\%$ compared to 7.3 $\%$ $\pm$ 2.3 $\%$ for the re--binned KH and OSIRIS
giants.

Our measurements of the CO and H$_2$O strengths for the re--binned KH spectra
correlate very well with
the the published KH indices. We find that the
CO index of KH, CO$_{KH}$, is related to our CO strength values derived
from the re--binned KH spectra, CO$_{rebin}$, as follows:
CO$_{KH}$ = $a$ + ($b$ $\times$ CO$_{rebin}$), with
$a$ = 8.90 $\pm$ 1.31 and $b$ = 1.27 $\pm$ 0.07.
Our H$_2$O strength measurements also agree well with those of KH,
with a mean difference between our measurements of the re-binned KH
spectra and the KH H$_2$O indices of $-0.4$ $\pm$ 0.4 \% (comparing
before extinction correction since KH made no correction).
We find, however, that the ``Na'' and ``Ca'' indices of KH are systematically
underestimated for stars with strong H$_2$O absorption,
such as BK Vir and SW Vir.
This is because KH adopted a continuum for the ``Na'' and ``Ca'' indices
which was interpolated linearly between two widely separated regions
in the blue half of the $K$ band, and this continuum is therefore
affected by the amount of H$_2$O absorption.
Our ``Na'' and ``Ca'' equivalent widths are derived using a local continuum,
and therefore are not as sensitive to the amount of H$_2$O absorption
(the presence of many weak \h2o \ lines might affect the overall level
of the continuum, Wallace \& Hinkle 1996).
We find good agreement of our ``Na'' and ``Ca'' equivalent widths, measured
from the original Terndrup et al. (1991) spectra, with the ``Na'' and ``Ca''
equivalent widths published by Terndrup et al. (1991).
All absorption measurements were computed using the LINER spectral
analysis program in use at Ohio State.

\section{RESULTS}

\subsection{Comparison with Previous GC Spectra}

Spectra for some of the GC cool stars in this paper
have been presented previously (LRT;
S87; Rieke et al. 1989; Krabbe et al. 1995). The OSIRIS spectra
presented in Figure~\ref{spect} are higher spatial resolution than
LRT, S87, and  Rieke et al. (1989), and similar to Krabbe et al. (1995).
Analysis of our $K-$band images (Paper I) shows that many of the
spectra presented in earlier, lower angular resolution work must have been
contaminated by neighboring stars
(particularly sources IRS 11, 12, 19, 22, 23, and 24).
For example, there is a star $\sim$ 1.8$''$ SW of IRS 11
and only 0.5 mag fainter (at $K$)
which must have contaminated the S87 IRS 11 spectrum (3.8$''$ beam
diameter) and also that of LRT (8$''$ beam diameter). There is
also a star 0.84 mag fainter than IRS 23 which was likely in the
beam of LRT and possibly S87 ($\sim$ 2.5$''$ to the NW).
These are probably
the worst cases for contamination of
the previously published spectra (see Table 1, Paper I),
although all the
sources in LRT, S87, and Rieke et al. (1989) must have had some contamination.
Such contamination can affect the measured absorption strengths of
the GC stars and lead to different estimates of stellar spectral
types for the GC stars.
Of the three stars for which we have analyzed both
OSIRIS and S87 re--binned data (IRS 11, 19, and 22), only IRS 11 shows a
statistically significant difference
in absorption strengths. Therefore, we adopt the OSIRIS values of the
absorption strengths for this star. For the remaining two stars (IRS 19
and 22) we adopt an average of the OSIRIS and S87 measurements.

When comparing the present data to previous work, there is
only one source upon which there has been disagreement over
the most basic spectral characteristics.
Krabbe et al. (1991,1995)
suggest that IRS 9 is likely a He I emission--line star, or
other hot star, based
on narrow--band imaging and a spectrum of He I (2.06 \mic) emission, but our
spectrum clearly shows IRS 9 to be a late type star based on
strong CO absorption at 2.3 \mic \ (Figure~\ref{spect}).
The Krabbe et al. (1995) spectrum of IRS 9 does not include the
2.3 \mic \ region.
Tamblyn et al. (1996) also make the cool star identification
based on narrow band imaging.
We note that there is small
residual Br$\gamma$ emission in this
source after background subtraction in our spectrum (possibly related
to this cool star; see below).

\subsection{Comparison to Disk and Bulge Stars}

\subsubsection{CO \& \h2o}

Figure~\ref{h2o} shows a comparison of measured CO vs. H$_2$O for the
GC and comparison stars.
The bulge giants, disk giants, and LPVs show a correlation
between CO and H$_2$O.
A small sample of CO and \h2o \ data for disk supergiants is also shown
in Figure~\ref{h2o}. These data show a different relation between
CO and H$_2$O than the giants and LPVs.

CO absorption strength increases
with decreasing effective temperature ($T_{\rm eff}$), decreasing gravity,
increasing [C/H], and increasing microturbulence (Baldwin et al. 1973;
McWilliam \& Lambert 1984). This last parameter increases with
increasing luminosity (McWilliam \& Lambert 1984; McWilliam \& Rich 1994).
For a star evolving up the giant branch
these factors conspire to produce the strong observed
correlation of increasing CO absorption strength
with increasing $J-K$, where increasing $J-K$
is due to decreasing \teff \ (McWilliam \& Lambert 1984).
This correlation holds for giants, on average, to very red $J-K$ but exhibits
large scatter in CO for the reddest stars ($J-K$ \apge 1.2--1.3 mag). This is
true of disk stars (McWilliam \& Lambert 1984) and bulge stars (FW87).
Much of the increased scatter in both disk and bulge star samples is due to
the presence of LPVs in the M giant samples (McWilliam \& Lambert 1984; FW87).

\h2o \ absorption strength also increases with decreasing \teff, but
decreases with increasing luminosity (Persson et al. 1977;
Aaronson et al. 1978; KH; Wallace \& Hinkle 1996).
A large increase in luminosity (accompanied by a decrease in gravity and
increase in
microturbulent velocity) leads to significantly higher CO absorption.
It is this contrasting luminosity dependence in CO and \h2o \ which
leads to the separation of giants and supergiants in
Figure~\ref{h2o}.

FW87 and Terndrup et al. (1991) used $(J-K)_\circ$ as a temperature
indicator for the BW and disk stars. Because of the much larger interstellar
extinction towards the GC, we cannot use the intrinsic colors of the
stars as a temperature indicator. We will use the CO strength
as a rough indicator of temperature (KH), while keeping in mind CO
also depends on luminosity.

The GC and bulge stars have generally stronger \h2o \
at a given CO
strength than the disk stars, and similar \h2o \ as the LPVs.
The H$_2$O measure
is sensitive to the derived \ak.
If the derived extinction for a
GC star is too high, the H$_2$O strength in this diagram is underestimated.
We have included in Figure~\ref{h2o} only those GC stars with $A_K$
determined from two or more infrared colors (Paper I).

The $A_K$ values applied here (Paper I)
employ the interstellar extinction curve
of Mathis (1990). This curve describes a power--law dependence of
extinction on the wavelength raised to the $-$1.7 power (an average
of recent determinations, Mathis 1990).
Other curves give slightly different powers.
The interstellar extinction curve
of Rieke \& Lebofsky (1985), for example, is well fit by a power$-$law
with exponent equal to $-$1.6. The law adopted by S87 (van de Hulst No.
15) is represented as a power$-$law
with exponent equal to $-$1.9.
The ``flatter'' law of Rieke \& Lebofsky (1985) would
result in $A_K$ about 10 $\%$ higher for
the same observed colors and assumed intrinsic colors.
However, the effect of a higher \ak \ (to make the derived \h2o \ absorption
smaller) is nearly cancelled by the effect of de--reddening the
spectrum with the ``flatter'' extinction law.
Using the Rieke \& Lebofsky
law to de--redden the spectrum of a GC star
and to derive \ak \ would
result in a measured \h2o \ value only $\sim$ 1.5 $\%$ less;
i.e, a GC star with \h2o \ of 12.0
$\%$ would shift to $\sim$ 10.5 $\%$ if we adopted the Rieke \& Lebofsky
extinction law. Similarly, if we adopted the same law as S87, the \h2o
\ would be $\sim$ 1.5$\%$ greater.

The CO and H$_2$O absorption strengths are also expected to change with
metallicity.  Observations of the integrated light of globular clusters
(Aaronson et al. 1978) and of individual stars at fixed V-K in globular
clusters and open clusters (Frogel et al. 1983; Houdashelt et al. 1992) show
that the CO absorption strength increases with increasing metallicity.
The H$_2$O absorption strength may also increase with increasing metallicity
(Aaronson et al. 1978).  The strong CO absorption in bulge stars has been
attributed to high metallicity (FW87; Terndrup et al. 1991) but
McWilliam \& Rich (1994) suggest that changes in surface gravity,
microturbulence, or the $^{12}$C/$^{13}$C ratio could also be important.

\subsubsection{CO, ``Na,'' \& ``Ca''}

KH identified Na and Ca as two strong atomic features in the $K-$band
for a large range of dwarves, giants, and supergiants. Recent higher
spectral resolution data indicates that the situation is more complex for
M supergiants and late M giants.
The high resolution spectra ($R \geq$ 45,000) of M giants and M supergiants
presented by Wallace \& Hinkle (1996)
show that
Sc I contributes as much or more to the total equivalent width
of both our ``Na'' and ``Ca'' measures.
Other significant contributors are Ti I, Si I, and V I, to ``Na'',
and Ti I and Fe I to ``Ca'' (Wallace \& Hinkle 1996).
Such contamination is confirmed for IRS 7;
a high resolution ($R$ $=$ 40,000) spectrum of IRS 7
in the ``Na'' feature shows
Sc is likely the largest single atomic contributor to
``Na'' in our spectrum (Carr et al. 1996$a,b$).
This strong Sc absorption complicates our interpretation
of the ``Na'' enhancement we observe in IRS 7 relative to
disk supergiants (see \S~4 below).

The GC stars and the BW stars both
appear to have stronger ``Na'' and ``Ca'' absorption strengths than the
field M giants and M supergiants (Figure~\ref{naca} and \ref{naca2}).
Note that there is more scatter for the ``Na'' measurement than for
``Ca'' in Figure~\ref{naca2}. It is not clear what causes this difference.
It is possible that our local continuum is more affected by the many
absorption lines near ``Na'' than for ``Ca'' in the GC
stars. There is a telluric absorption feature inside our ``Na'' bandpass;
variations in the correction of this feature could lead to differences
in derived absorption strength.
There may be small residual [Fe III] 2.217 \mic \ emission (Lutz et al.
1993) in some GC star spectra from incomplete background subtraction
which could have affected our continuum
placement as well. This latter possibility does not affect our
conclusion that the ``Na'' absorption is enhanced. Many of the GC stars with
strong ``Na'' absorption strength are not in regions where diffuse [Fe III]
emission is strong, while IRS 20, in a region of strong [Fe III]
emission, has a relatively small measured ``Na'' absorption strength.

Both ``atomic'' absorption
features may also be affected by molecular absorption.
There are many lines due to CN identified by
Wallace \& Hinkle (1996) in our
``Na'' and ``Ca'' bandpasses. Higher resolution
spectra of IRS 7 (S87; Carr et al. 1996$a,b$) suggest these are
important contributors as well.
Model results for giant and supergiant M stars show that
a substantial fraction of the total
equivalent width of our ``Na'' and ``Ca'' equivalent widths could be due to
CN absorption
(Carr et al. 1996$a,b$). The effect will also depend on how
CN affects the nearby continuum, but it is possible that the atomic
features speak as much to CN (and hence CNO processing) as to
atomic abundances.
Carr et al. (1996$a,b$) show that CO is weaker in IRS 7 compared to the
M2 I $\alpha$ Ori and that CN is stronger.
This is consistent with our finding that IRS 7 has weaker CO
strength than $\mu$ Cep (M2 Ia, see Table 1) but stronger ``Na'' and
``Ca;'' see Table 1 and Figures~\ref{naca} and \ref{naca2}.

\section{Discussion}

\subsection{AGB Stars and Supergiants}

The cool GC stars we have observed are more
luminous than the tip of the first ascent red giant branch, and
therefore
must be either AGB stars or supergiant stars.
We would like to distinguish between AGB stars and supergiants
because each traces different epochs of star formation in the GC
(Haller \& Rieke 1989; Haller 1992; Krabbe et al. 1995).
It is important to separate individual stars
between older star formation epochs
and more recent ones since the relative numbers
of red and blue supergiants present may be important in constraining
starburst models in the GC (Krabbe et al. 1995).
LRT and S87 made earlier attempts to distinguish between M giants
and supergiants. Here, we re-address this question with our higher
angular resolution data (both the spectra presented here, and the
photometry in Paper I) and in the context that some of the stars
may be AGB stars, in particular, LPVs.

\subsubsection{Definitions and Observed Characteristics}

In this discussion, we will use the following definitions for AGB stars
and M type supergiants taken from Jones et al. (1983). By AGB stars,
we mean those intermediate and low mass
stars (initial masses, M \aple 9--10 M$_\odot$) with degenerate C/O cores
which are producing their
luminosity through helium shell and hydrogen shell fusion.
The brightest AGB stars are thermally pulsing variables
(see Iben \& Renzini 1983 and Wood 1990 for reviews of AGB stars) known as
Miras and OH/IR stars. Generally, we will call such stars LPVs or AGB stars.
AGB stars are generally believed to obey a core-mass vs. luminosity
relationship (Paczy\'{n}ski 1970) which limits their maximum
luminosity to $M_{\rm bol}$ $=$ $-$7.0 mag.
Recent modeling of a 7 M$_\odot$ AGB star
suggests slightly higher luminosities ($M_{\rm bol}$ $<$ $-$7.2) are
possible (Bl\"{o}cker \& Sch\"{o}nberner 1991).
Supergiants are those stars still burning helium (or carbon)
in their cores. They will
have initial masses generally greater than 10 M$_\odot$.
M supergiants can
reach bolometric luminosities up to $M_{\rm bol}$ \aple $-$9.5 mag, but
some have $M_{\rm bol}$ less luminous than $-$6.0 mag (Humphreys 1978;
Figure~\ref{cmd}).

LPVs should be regular pulsators with large amplitude variability and
long periods. This is strong incentive for continued and more regular
variability searches in the GC.
While variable, M supergiants do not appear to behave like
normal Miras; their variations are typically of smaller amplitude
(\aple 0.5 mag at $K$) and irregular (Harvey et al. 1974;
EFH; Jones et al. 1988).
Jones et al. (1988) do discuss a small subset of
luminous OH/IR stars which may
``masquerade'' as LPVs, but which are really more massive supergiants.
However, these stars are more luminous than the AGB limit.
There also exists one optically visible
M supergiant (VX Sgr, a strong OH emitter)
that may be related to these luminous OH/IR stars (Jones et al. 1988)
but in no way appears to be typical of M supergiants or Miras.
This star has semi--regular photometric variations (Harvey et al. 1974)
and moderate H$_2$O absorption (Hyland et al. 1972; Jones et al. 1988).
To our knowledge, it is the one known star with primarily
Mira like characteristics and a supergiant classification (based
on spectral characteristics, Lockwood \& Wing 1982).
Elias et al. (1980) argue that VX Sgr may be similar to the
Small Magellanic Cloud (SMC) large amplitude (LA)
variables, i.e., it is in fact an AGB star. We note that
the estimated distance
to VX Sgr (and hence luminosity) may be quite uncertain, ranging from
$\sim$ 250 pc (Celis 1995) to 1500 pc (Lockwood \& Wing 1982).

The $H-K$ colors of the Large Magellanic Cloud (LMC), BW M giants,
and Sgr I LPVs in Figure~\ref{cmd}
have independent measurements of $A_K$ (the $A_K$ is not derived
from the observed colors of these stars). Their generally redder colors
result from a combination of circumstellar emission,
circumstellar extinction, and
photospheric colors (Feast et al. 1982;
Whitelock et al. 1986; Gaylard et al. 1989;
Glass et al. 1995; Zijlstra et al. 1996; see also
the discussion on the Sgr I LPVs in Appendix 1).
This is assumed
to be the case for the much redder IRAS LPVs as well. The reddening
vector in Figure~\ref{cmd} suggests that the $H-K$ colors of the IRAS
sources might be
dominated by circumstellar extinction.

The $K-$band spectra
of known Mira variable stars (e.g., $o$ Cet, R Hya, R Cas,
and R Leo; Johnson $\&$ M\'{e}ndez 1970; Hyland et al. 1972;
Merrill \& Stein 1976; Strecker et al. 1978)
are characterized by strong CO and H$_2$O absorption.
The H$_2$O absorption produces a large
depression in the continuum which increases in strength
toward the blue end of the $K-$band (Figure~\ref{spect})
which cannot be accounted for by interstellar extinction.
Narrow--band photometry (Frogel 1983; EFH) and low
resolution spectroscopy (Jones et al. 1988) have also clearly
associated strong H$_2$O absorption with LPVs.
Miras may also exhibit
Br$\gamma$ emission, the appearance of which may be variable
(Johnson $\&$ M\'{e}ndez 1970).

\subsubsection{Color--Magnitude Diagram}

Figure~\ref{cmd} presents the de--reddened color--magnitude diagram (CMD)
for the GC cool stars
and compares it to other well--studied populations of supergiants, giants, and
LPVs. The GC stars are compared separately
to supergiants and AGB stars.
Each of the data sets presented in Figure~\ref{cmd} is described in
Appendix 1. Here we mention several important points.
The GC stars have $A_K$ from Paper I which were determined by assuming
intrinsic colors. The $A_K$ may be overestimated (\aple 0.5
mag) for some
GC stars (see the discussion in Appendix 1).
The Milky Way supergiants have distances
derived from the OB stars in their individual associations.

IRS~7 is the only M star in our spectroscopic
sample of the central $\sim$ 5 pc that must be a
massive supergiant based on its luminosity.
IRS 7 has $M_{\rm bol}$ $=$ $-$9.0 mag (BC$_K$ $=$ 2.6 mag for M2 I, EFH;
see also Appendix 1).
The remainder of the GC stars are below
the theoretical upper limit on bolometric
luminosity ($M_{\rm bol} =$ $-$7.0) for
lower mass AGB stars if we apply the BC$_K$ (2.9$-$3.2 mag)
derived from known LPVs (see Appendix 1).
More than half
of GC stars have relatively faint $M_K$ (\apge $-$8.0 mag).
These stars have $M_K$ fainter than the faintest
supergiant in our comparison sample.
However, the Milky Way supergiants do overlap with the
LPVs in the range
$-$10.4 \aple $M_K$ \aple $-$8.1 mag ($-$7.8 \aple $M_{\rm bol}$ \aple $-$5.5
mag) and some of the GC stars lie in this range. Clearly
$M_K$ cannot be used to unambiguously separate all the GC stars.
At this point we can bin the GC stars in Table 2 into three groups based
on $M_K$ (assuming, for the moment, they are all M stars):
1.) definite supergiants, IRS 7;
2.) M giant or AGB star, IRS 12S, 20, OSU C1, C2, and C4;
3.) supergiant or luminous AGB star, IRS 1NE, 1SE, 2, 9, 11, 12N, 14NE,
19, 22, 23, 24, 28 and OSU C3.
We do not have spectra of the remaining
stars in Table 2, but all have $M_K$ $\geq$ $-$8.0 mag and
cool star identifications based on published spectra (Krabbe et al.
1995; Figer 1995).

\subsection{AGB and LPV Stars in the GC: CO, \h2o, \& Variability}

We have spectra of  13 GC stars which have $M_K$ consistent with either
a supergiant or AGB star classification. Seven of these stars (IRS 9,
11, 12N, 14NE, 23, 24, and 28) have strong \h2o \ absorption characteristic
of LPVs. IRS 12S, with $M_K$ corresponding to an AGB star, also
has strong
\h2o \ characteristic of LPVs.
These GC stars have
CO strength similar to the known LPVs (Miras) in Table 2 with the
exception of IRS 23. Inspection of Figure~\ref{spect} suggests the
continuum position of IRS 23 may be affected by additional
absorption which depresses it relative to the other stars (also recall
that more than one star may contribute to the S87 spectrum of this
star).

In addition to having strong \h2o \ like the disk LPVs, IRS 9, 12N, 24,
and 28 have
recently been identified as
photometric variables at $K$ (Haller 1992; Tamura et al. 1994, 1996)
and $J$ (IRS 9 and 12N, Paper I). IRS 23 is just below Haller's (1992)
three sigma cut--off for variability at $K$.
The spectral
characteristics of IRS 9, 12N, and 28 coupled with the photometric variability,
strongly support the suggestion by Tamura et al. (1996) that
these stars are LPVs like the Miras. A similar case now exists for
IRS 24 and most likely IRS 23.
Recent searches for variability in the GC
(Haller \& Rieke 1989; Haller 1992; Tamura et al. 1994, 1996),
while ground breaking, are only a first
step at characterizing the stellar variability in the GC.

IRS 24 has also been identified with an
\h2o \ maser (Levine et al. 1995) and an OH maser (Sjouwerman \& van
Langevelde 1996).
We believe these maser identifications strengthen the case
for IRS 24 being an LPV, but see Levine et al. (1995) for a different
interpretation. IRS 24 and additional OH
masers in our field are discussed in Appendix 2.

Two of the GC stars with $M_K$ implying they are either supergiant
or AGB stars (IRS 19 and 22) have strong CO,
luminous $M_K$, and weak \h2o. We conclude that these stars
are likely supergiants.
We note that
Blum et al. (1996, Paper I) find IRS 7 to
be variable but with variation ($\sim$
0.3 mag at $K$)
which is consistent with supergiants (Harvey et al. 1974; EFH).
It is the combination of weak \h2o, strong CO, and $M_K$ which leads
to the supergiant classifications.

This leaves four GC stars (IRS 1SE, 1NE,
2,and OSU C3) with combinations of CO, \h2o, $M_K$,
and variability that do not firmly establish a classification. These
four stars may be AGB stars (M7 or later) or K5--M0 supergiants.

The classifications (or lack thereof) for all the GC stars
for which we have spectra are summarized in Table 2. Within the
uncertainties posed by the ambiguous group, we can now compare the
CO and atomic line absorption strengths of the GC stars with
similar measures from stars of known populations to further explore
the GC star properties.

\subsection{Absorption Strengths: GC vs. Bulge and Disk Stars}

\subsubsection{CO}

Figure~\ref{h2o} suggests that the GC stars with supergiant
classifications (IRS 7, 19, and 22) all have normal CO strengths for M 0--2
supergiants. The GC stars with AGB or LPV classifications have CO
which range from about 14 $\%$ to 23 $\%$. On average, the
GC AGB/LPV stars appear to have stronger CO than the latest bulge and disk
M giants. This is most likely due to the fact that the GC stars
represent the  coolest, most luminous stars on the AGB; investigation
of the CO {\it abundance} and of the GC population will be better
addressed with larger samples of stars which span much earlier
spectral types (e.g. FW87). Note that
none of the GC AGB/LPVs has CO absorption which is significantly
stronger than the CO absorption in the Mira R Cas.

If we assume that the GC stars have CO strengths similar to the disk
stars of their respective classes (e.g. supergiant or AGB) we can assign
spectral types to them and, hence, obtain an estimate of \teff \ for
each star. We estimate spectral types to the nearest integer sub--type
by comparing the GC star CO strength to the representative members
of its class in the disk star sample of Table 1. This results in the
spectral types shown in Table 1. We adopt the spectral type vs. \teff \
calibration of Dyck et al. (1996)
for the AGB stars.
For the LPVs, it does not appear that optical
spectral type is well correlated with \teff.
Ridgway et al. (1992)
showed that $o$ Cet has much lower \teff \ than non$-$Mira
M giants of the same optical spectral type.
Therefore, we estimate \teff \ in a different way for the LPVs.
The luminosity of LPVs appears to be tightly correlated with mass
(Vassiliadis \& Wood 1993; Jones et al. 1994).
The evolutionary models of Vassiliadis \& Wood (1993) suggest masses
based on our $M_{\rm bol}$ estimates for the GC stars (see Table 2)
which lead to
estimates of \teff \ (from the Vassiliadis \& Wood models) that are
in close agreement with the derived \teff \ for $o$ Cet (2300 K) from
Ridgway et al. (1992); see the discussion on LPV masses in \S~4.4.

We adopt the spectral type vs. \teff \
calibration of Johnson (1966) for the supergiants; the Johnson scale
is consistent with the results of Dyck et al. (1996).
The effective temperature of $\alpha$ Ori (M2 I)
determined from its angular diameter (Dyck et al. 1992, 1996) is in excellent
agreement with the Johnson (1966) value for M2 I (3600 K).
The \teff \ for GC stars is given in Table 2.

By using the above CO (spectral type) vs. \teff \ calibration we have
assumed the GC stars have similar metallicity and CO abundance as the
disk stars. This may not be the case. Detailed metallicity and abundance
analyses are just now being completed for some GC stars (Carr et al.
1996$a,b$). If it is found that the GC metallicity scale is much different
than the disk stars, we may need to adjust our spectral type and \teff \
assignments.

\subsubsection{``Na'' \& ``Ca''}

Figure~\ref{naca} is strongly affected by
$T_{\rm eff}$ and by luminosity. However, the luminosity
effect is almost entirely in the CO strength which separates
the disk and bulge giants from the
disk supergiants.
The absorption strength of the ``Ca'' and ``Na'' lines
is observed to increase with decreasing $T_{\rm eff}$ (KH;
Terndrup et al. 1991; Ram\'{\i}rez et al. 1996; all three
analyses are affected by the contamination discussed above).
The high resolution data of Wallace \& Hinkle (1996) also
show this trend for the various contributors to ``Na'' and
``Ca'' which were identified in \S~3.
Figure~\ref{naca} and \ref{naca2} (disk giants and supergiants) each show a
correlation of ``Na'' and ``Ca'' with CO
where CO now represents, for a fixed luminosity class,
a measure of $T_{\rm eff}$.

Figure~\ref{naca} suggests that the GC and BW stars have higher ``Na'' and
``Ca'' than the disk stars of similar CO strength.
For the bulge stars, Terndrup et al. (1991) considered
a range of temperatures for both disk and bulge giants and showed
that, on average, the
``Na'' and ``Ca'' strengths for a given temperature were higher for bulge
stars.
This effect can be seen in Figures~\ref{naca} and ~\ref{naca2} for the
bulge M giants. Consider
stars with CO greater than 15 $\%$, which corresponds to
the $T_{\rm eff}$ range (actually $J-K$ $\geq$ 1.00 mag)
for the Terndrup et al. (1991) bulge and disk giants
analyzed here. The bulge stars have mean
``Na'' plus ``Ca'' which is higher than the disk giants, 12.48 $\pm$
0.67 \AA \ compared to 10.05 $\pm$ 0.31 \AA. The enhancement also
holds for ``Na'' and ``Ca'' individually. The range of $M_K$ for the
BW giants suggests some may be on the AGB while others may be first
ascent giants. This is likely true of our disk star sample too. Therefore,
our comparison could be affected by differences in ``Na'' and
``Ca'' absorption resulting from these different evolutionary states.

If we consider the GC AGB/LPV stars
in the same range of CO strength as above, we
find that ``Na'' plus ``Ca'' (12.29 $\pm$ 0.82 \AA)
is also stronger than for the disk stars, and similar to the value
for the bulge stars (``Na'' and ``Ca''  are stronger individually too).
As for the disk and bulge giant comparison,
we caution, that we are primarily
comparing disk giants to GC AGB stars since our disk sample only
includes one LPV
(re--binned high resolution spectrum
of $o$ Cet from Wallace \& Hinkle 1996). We did not attempt
``Na'' and ``Ca'' measurements of the Johnson \& M\'{e}ndez (1970) LPVs
due to the lower spectral sampling and poorer
signal--to--noise. A larger sample of cool LPVs may show
average ``Na'' and ``Ca'' which are stronger than the other disk giants
in Figure~\ref{naca}.

Terndrup et al. (1991) argued that the enhanced absorption strengths
of ``Na'' and ``Ca'' in BW reflected enhanced {\it abundances}
relative to disk giants. Their measurements were susceptible to the same
contamination problems we discussed above, so it is not clear that
the BW stars actually have enhanced Na or Ca, although this is, of
course, not ruled out since high resolution
spectra have not been obtained in the $K-$band for the
BW giants. McWilliam \& Rich (1994), in detailed
abundance analyses using high resolution optical spectra, found that [Ca/Fe] in
a sample of bulge K giants (11 stars) was similar to that expected
for disk or halo stars of a given [Fe/H], while Na might be
over--abundant in several of their stars.

The GC supergiants show a clear enhancement of ``Na'' and ``Ca''
absorption over the disk supergiants,
as pointed out for IRS 7 by S87. The mean value of ``Na''
plus ``Ca'' for the three GC supergiants is 13.4 $\pm$ 1.0 \AA. By
contrast, the strongest value we find for a disk
supergiant is 11.9 \AA \ for SAO 11969 (M3 I), and the remainder of the
disk supergiants have ``Na'' plus ``Ca'' less than 11.5 \AA.
Note that this enhancement is seen in ``Na'' and ``Ca''
individually  as well (Figure~\ref{naca2}).
IRS 7, clearly a M1--2 supergiant,
has ``Na'' plus ``Ca'' 2.5 $\pm$ 0.7 \AA \ greater than SAO~11969.
A high resolution spectrum of IRS 7 suggests
that Sc I, which contributes significantly to the ``Na'' equivalent width,
appears enhanced relative to $\alpha$ Ori (M2 I)
(Carr et al. 1996$a,b$; see discussion in \S~3).
Sc is an iron-peak element whose abundance should follow that of iron
(Wheeler, Sneden, \& Truran 1989).
However, IRS 7 and $\alpha$ Ori exhibit essentially the
same [Fe/H] as derived from infrared spectra (the value is
nearly solar, Carr et al. 1996$a,b$). A current lack of laboratory data for
the hyperfine splitting in Sc makes it difficult to accurately
model its line strengths and derive a precise abundance.
An interesting possibility is the enhancement of the
Sc abundance by mixing to the surface the products of mild $s-$processing.
Smith \& Lambert (1987)
predict that the Sc abundance could be significantly enhanced (factor of three)
by mild $s-$processed material mixed to the surface in late--type stars.
This would require more mixing of processed material than is seen
in similar disk stars (e.g. $\alpha$ Ori), assuming the same process
occurs in the stellar interiors of disk and GC stars. While
helping to solve our Sc abundance problem, we would
then need an explanation for the increased mixing in GC stars.

At present, therefore, we do not know whether
the strong Sc absorption in IRS 7 (which may also be a significant
contributor to the
strong ``Na'' and ``Ca'' in other GC stars) is due to selective enrichment
of Sc (by mild
s--processing or some other mechanism) or whether Sc is particularly
sensitive to some parameter in the model atmospheres, such as
surface gravity or microturbulence, which differs slightly between
IRS 7 and $\alpha$ Ori.
We plan to make high spectral resolution measurements
of the ``Na'' and ``Ca'' features to determine the underlying causes of the
demonstrated enhancement in absorption strengths.
We plan to explore Sc enhancement as well as the effect of the
remaining atomic (Ti, V, Si, and Fe) and molecular (CN) lines
which contribute to the ``Na'' and `Ca'' strengths (see \S~3).

\subsection{Masses and Ages}

We may estimate the masses and ages of the GC stars
by comparing estimates of their bolometric luminosities and
effective temperatures (Table 2)
to stellar evolution models.
The M1 I classification for IRS 7 (Table 1) is consistent with
previous determinations (LRT, S87)
and gives $T_{\rm eff}$ $=$ 3600 K for our adopted spectral type
calibration (\S~4.3.1).
The detailed analysis of weaker CO lines from high
resolution spectra (Carr et al. 1996$a,b$)
is consistent with this temperature. Taking 3600 K and using the
derived value of $M_{\rm bol}$ ($-$9.0 mag, BC$_K$ $=$ 2.6 mag) suggests
an initial mass of $\sim$ 20--25 M$_\odot$ and age of 7--9 Myr for the
Z $=$ 0.02 evolutionary tracks of Schaller et al. (1992).
The solar metallicity tracks were chosen based on the Carr et al. (1996$a,b$)
finding that [Fe/H] in IRS 7 is nearly solar.
Similarly,
the values of $M_{\rm bol}$ and \teff \ given in Table 2 suggest masses
of 12$-$15 M$_\odot$ for IRS 19 and 9$-$12 M$_\odot$ for IRS 22,
again for Schaller et al. (1992) Z $=$ 0.02 tracks.
Corresponding ages are $\sim$ 12$-$18 Myr and 18$-$29 Myr.

The situation for the candidate LPVs (IRS 9, 12N, 23, 24, and 28)
is more speculative.
Few models exist for evolution near the top of the AGB; such models depend
on empirically determined mass--loss rates (Wood 1990).
Comparison to the Z $=$ 0.016 (largest Z for which models were computed) models
of Vassiliadis \& Wood (1993) will allow for a rough estimate.
Using the BC$_K$ derived from the Sgr I LPVs (3.2 mag, Appendix 1),
the GC LPV candidates have
$-$5.7 \apge $M_{\rm bol}$ \apge $-$6.5 mag.
The evolutionary models of Vassiliadis \& Wood (1993) predict the
relatively narrow mass range of $\sim$ 4--5 M$_\odot$ for (Mira--like)
LPVs in this luminosity range.
Adopting these masses for the GC LPVs and using their derived $M_{\rm
bol}$ gives an estimated \teff \ of $\sim$ 2600 K, also from the
models of Vassiliadis \& Wood
(1993). This agrees well with the value mentioned above for $o$ Cet, the
proto--type Mira.
The corresponding ages from the Vassiliadis \& Wood models are
120--200 Myr, similar to the 107--190 Myr ages of 4--5 M$_\odot$ stars from
the Schaller et al. (1992) models.

The remaining AGB stars in Table 2 have lower $M_K$ and thus $M_{\rm
bol}$. The CO strengths and $M_K$ are consistent with a late M
classification, so their BC$_K$ are similar to the BC$_K$ of the
Sgr I LPVs (FW87, Glass et al. 1995).
For simplicity, we use the same BC$_K$ (3.2 mag, Appendix 1).
We estimate, by
comparison to the Schaller et al. (1993) tracks, that the remaining
AGB stars have maximum initial masses of $\sim$ $3-4$ M$_\odot$,
corresponding to minimum ages of $\sim$ $190-440$ Myr.
The tracks of Vassiliadis \& Wood (1993) for AGB stars suggest that the
brightest GC AGB stars, IRS 11 and 14NE, must have minimum initial masses
$\sim$ 2 M$_\odot$ in order to reach their observed luminosities. This
corresponds to maximum ages of $\sim$ 1.6 Gyr.
Similarly, IRS 12S, OSU C1, C2, and C4 have minimum initial masses of
$\sim$ 1 M$_\odot$ and maximum ages of $\sim$ 12 Gyr.
Extrapolating the results of Vassiliadis \& Wood (1993) to the
luminosity of IRS 20 suggests a minimum mass of about 0.8 M$_\odot$
for this star. The corresponding age from the Schaller et al. (1993)
Z $=$ 0.02 tracks is $\sim$ 25 Gyr.
Since this is longer than
the age of the universe (Sandage 1988; Freedman et al. 1994), IRS 20
is likely an AGB star which is more massive than 0.8 M$_\odot$ but which
is still evolving up the AGB and has not reached its maximum luminosity
on the AGB yet.

We have been unable to distinguish between AGB stars or supergiants for
four of the GC stars: IRS 1NE, 1SE, 2, and OSU C3.
If these stars are AGB stars they would have
M $<$ 3--5 M$_\odot$ and age $>$ 120--440 Myr.
The minimum masses for this case would be approximately
1,2,2, and 4 M$_\odot$ for IRS 1SE, 1NE, OSU C3, and IRS 2,
respectively, with
corresponding maximum age of $\sim$ 200 -- 12000 Myr.
However, using BC$_K$ and \teff \ as for the supergiants above,
IRS 1NE, 1SE and OSU C3 could have initial masses as high as 9 M$_\odot$
and ages $\sim$ 29 Myr. Similarly, we
would estimate M $\sim$ 12 M$_\odot$ and age $\sim$ 16 Myr for IRS 2
if it is a supergiant.

The estimates of $M_{\rm bol}$, \teff, mass, and age are summarized in
Table 2. The difference in age for these luminous M stars suggests that
there have been multiple, recent epochs
of star formation in the GC. Star formation models have been computed in
the GC (Tamblyn \& Rieke 1993; Krabbe et al. 1995). These models rule
out ``continuous'' star formation on the grounds that this mode would produce
many more red supergiants relative to the known blue supergiants (i.e.
the emission--line stars; see \S~1). A picture of
star formation in the GC which is consistent with the results summarized
in Table 2 and the aforementioned model results is one in which
small bursts have occurred
in a quasi--periodic fashion at different locations
leading to the mix of luminous
stars we presently observe in the central $\sim$ 4$-$5 pc. In this case,
IRS 7 would be one of the older stars in the
3--7 Myr burst model of Krabbe et al. (1995) which
also accounts for the large number of massive emission--line stars
in the central parsec. The other GC stars result from earlier
star formation epochs, $\sim$ 10$-$30 Myr and \apge 100 Myr ago.

\section{SUMMARY}
We have compared the $M_K$, CO absorption, and \h2o \ absorption
for a sample of GC stars
to the same quantities derived from photometry and $K-$band spectra
of known populations
of supergiants, giants, and LPVs. The GC stars span the brightest
five magnitudes in the observed $K-$band luminosity function presented
by Blum et al. (1996, Paper I).
Of all the bright stars in the GC
identified as cool, based on $K-$band spectra, only IRS 7 must
be a younger supergiant. The remainder are consistent with less massive
M giants and LPVs. Based on bolometric corrections from the
known populations of LPVs, none of the
GC stars (except IRS 7) has $M_{\rm bol}$ which exceeds the
theoretical limit of $-7.0$ mag for AGB stars.
Some GC stars do have luminosities which overlap with the less
luminous Milky Way supergiants. Two of these (IRS 19 and 22)
exhibit CO and \h2o \ absorption
characteristic of supergiants; we therefore classify them as such.
All but four of the remaining GC stars for which we have spectra
are classified as AGB/LPV stars. Classification for four stars
remains ambiguous between AGB and supergiant.

The coolest, most luminous AGB stars are LPVs.
Our $K-$band spectra of four photometric variables
in the GC (IRS 9, 12N, 24, and 28)
and one possible variable (IRS 23)
show extreme \h2o \ absorption which is remarkably similar to
that exhibited by known galactic LPVs. Based on this similarity
and variability, we conclude that they are very likely
LPVs, as suggested for IRS 9, 12N, and 28 by Tamura et al. (1996).
A similar conclusion was reached by Sjouwerman \& van Langevelde (1996)
for IRS 24 based on its OH maser characteristics.
Evolutionary models suggest these LPVs
had initial masses of $\sim$ 4--5 M$_\odot$ and are roughly 100--200 Myr old.
This is in contrast to the M1--2 supergiant, IRS 7, for which we
estimate an initial
mass of 20--25 M$_\odot$ and age of 7--9 Myr.
The other GC supergiants may have masses in the range 9--15 M$_\odot$
and ages of 12--29 Myr.
We believe the luminous, cool stars in the GC are tracing multiple epochs
of star formation, perhaps as quasi--periodic bursts, over the
last 7--100 Myr.

Our analysis of the spectra for the cool stars
in the GC results in absorption strengths of CO, H$_2$O, ``Na'',
and ``Ca.'' The atomic measurements of ``Na'' and ``Ca''
from our low spectral resolution
data are shown to be severely contaminated by
other atomic species including Si, Fe, Ti, Sc, and V and also molecular
lines of CN.
The measurements for the four stars we classify as supergiants show a
clear enhancement in ``Na'' and ``Ca'' over disk supergiants.
This includes the well--known GC supergiant IRS 7 which shows a significant
enhancement relative to
the disk supergiant in our sample with the largest ``Na'' and
``Ca'' strength. One component contributing to the enhancement in IRS 7
is Sc, an iron--peak element which may be enhanced by mild
$s-$processing. In general, the cause of the enhancement remains a
puzzle which we plan to investigate further with high spectral
resolution observations.

The AGB/LPV stars in the GC also
have enhanced ``Na'' and ``Ca'' relative to disk giant stars. The measured
``Na'' and ``Ca'' strengths in these GC stars
are similar to those found in the coolest bulge M giants. Temperature
and luminosity effects make it difficult to determine abundance effects
in the GC AGB stars relative to disk stars.
We plan to make additional high
resolution measurements of AGB/LPV GC stars, like the supergiants,
in order to better understand
the underlying causes of the enhanced ``Na'' and ``Ca'' absorption
strengths.

This work was supported by National Science Foundation grants AST
90--16112, AST 91--15236, and AST 92--18449.
Support for this work was also provided by NASA through grant number
HF 01067.01 -- 94A from the Space Telescope Science Institute, which is
operated by the Association of Universities for Research in Astronomy,
Inc., under NASA contract NAS5--26555.
We thank D. Figer, D. Levine, M. Hanson, D. Terndrup, and
L. Wallace  for supplying us with spectra.
The Johnson \& M\'{e}ndez (1970) and KH
spectra presented and/or analyzed in this paper are available
from the Astronomical Data Center (ADC) which is maintained
at the NASA Goddard Space Flight Center. We are grateful to S. Kleinmann, D.
Hall, H. Johnson, and M. M\'{e}ndez for putting
these spectra on the ADC electronic database.
We thank J. Carr for very useful
discussions relating to the absorption strengths of molecular
and atomic species in M stars.
Once again, it is a pleasure to thank R. Pogge for his LINER program.
This research has made use of the SIMBAD database, operated at
CDS, Strasbourg, France.


\section{Appendix 1}
The photometry used in constructing Figure~\ref{cmd} is described herein.

{\it Galactic Center cool stars.} The GC photometry is described in
Paper I.
Here, we adopt a distance to the GC of 8 kpc (Reid 1993). The
extinction ($A_K$) was derived by {\it assuming} intrinsic colors for
the GC stars, so we cannot distinguish the spectral type of the
GC stars by color.
However, the spectra in Figure~\ref{spect} identify all the GC
stars as M stars based on their strong CO absorption (and/or H$_2$O;
e.g, IRS 23).
IRS 1SE and 2 may be an exceptions, but they would have to be
a late K supergiants if not an M giants (compare HR 8726 in Table 1).
The intrinsic colors for normal M giants and supergiants
do not vary a great deal; therefore,
adopting a set of intrinsic colors on the giant branch
leads to estimates for $A_K$ which are relatively
accurate.
In Paper I, stars with measured $J-H$ and $H-K$ had adopted values of
0.7 mag and 0.3 mag, respectively, for these colors from FW87 (these are
similar
to the intrinsic colors of M supergiants, EFH).
If only $H-K$ was available, the star was de--reddened to a mean
relation which resulted in a similar $H-K$.
It is unlikely that any of the GC stars has
intrinsic colors substantially
more blue than these. Therefore, it is unlikely that we
have substantially underestimated the intrinsic luminosity of any GC
star in
Figure~\ref{cmd}. For
example, an $H-K$ of 0.2 mag, instead of 0.3 mag, would result in $A_K$
and the $K$ luminosity being underestimated by 0.16 mag.

What if some of the GC stars are LPVs, not normal giants or
supergiants?
As we will see below, LPVs may have redder colors than we adopted in
Paper I for the GC stars.
This means any GC star which is actually an LPV is likely to have had
its intrinsic luminosity overestimated.
The $J-H$ and $H-K$ colors of the reddest LPVs in Glass et al.
(1995) are 1.10 and 0.61 mag, respectively. If these colors were adopted
for the GC LPV candidates (Table 2), the resulting $A_K$ would be 0.45
mag less and $M_K$ would therefore be 0.45 mag fainter.
Additionally, the color$-$color diagram for
stars in the GC field
(Paper I) suggests a minimum \ak \ of \apge 2.0 mag for any star likely
to be physically located near
the GC. This value can be used to estimate a minimum luminosity for any
GC star (Table 2).

We have plotted in Figure~\ref{cmd} all stars in the GC for which we have
obtained at least $H$ and $K$
photometry and for which cool star identifications
exist (Paper I). The absolute $K$ magnitudes ($R_\circ$ $=$ 8 kpc)
and $A_K$ from Paper I are listed in Table 2.

{\it Milky Way and LMC M supergiants.} The supergiant data were drawn
from the sample of EFH
which they used to derive intrinsic colors
and bolometric corrections
for Milky Way and LMC supergiants. The Milky Way data shown in
Figure~\ref{cmd}
are the subset of stars from the EFH sample with distances determined
by Humphreys (1978) from the OB stars in individual associations.
The majority of Milky Way supergiants plotted in
Figure~\ref{cmd} are luminosity class
Iab (22 stars), but luminosity classes range from Ib (four stars) to Ia
(five stars). The 31 confirmed LMC supergiants in EFH with infrared
photometry are nearly equally split among luminosity classes Ia (15
stars) and Iab (11 stars) with the
remainder having no luminosity class assigned.
EFH found that the bolometric correction for Milky Way supergiants is
nearly constant among sub--types
and that this value is quite similar to those determined
for supergiants in the LMC and SMC which are well known to have lower
metallicity than Milky Way stars (e.g, Dufour 1986 and references
therein). The full range of BC$_K$ for Milky Way,
LMC, and SMC M supergiants is 2.5--2.8 mag.

The $H-K$ and BC$_K$ of M supergiants which we have adopted (EFH) are
slightly smaller than the values given by Lee (1970). Lee's $H-K$
are up to 0.10 mag more red for type Ia supergiants and the BC$_K$
derived from Lee's $V-K$ and BC$_V$ are up to $\sim$ 0.3 mag larger;
i.e, for the same $M_{\rm bol}$, we would compute $M_K$ up to
0.3 mag brighter. This would tend to separate the supergiants more from
other stars in Figure~\ref{cmd}.

{\it Milky Way M giants.} The Milky Way M giants are plotted as the
solid line in Figure~\ref{cmd} which rises to $M_K$ $=$ $-$8.0 mag. The
line was obtained by using the relationship for absolute visual
magnitude vs. MK spectral type derived by Th\'{e} et al. (1990) in
combination with the $V-K$ and $H-K$
colors of Lee (1970) and Frogel et al. (1978).
The adopted $V-K$ are similar to the values given by Johnson (1966).
The colors for type M7 III are an average of those for BK Vir and SW
Vir (Frogel et al. 1978; Wisse 1981; Mermilliod 1987).

{\it Sgr I LPVs.} These are Mira variables observed in the optical
window Sgr I ($l, b =$1.5\deg, $-$2.7\deg) by Glass et al. (1995). A
distance of 8 kpc to the GC
(Reid 1993) was assumed to derive absolute $K$ magnitudes. The
data were de--reddened by $\sim$ 0.2 mag at $K$ (Glass et al. 1995).
Differences
between the two systems at $H$ and $K$ are unimportant
for this comparison: \aple 5 $\%$ at $K$ and \aple 1 $\%$ at
$H-K$ (see the transformation from SAAO to CTIO, Glass et al.
1995 and Carter 1990), so the Sgr I photometry is not
transformed on to the same system as EFH (CIT/CTIO).
Glass et al. show that these
stars are significantly redder than solar neighborhood M giants and
known solar neighborhood Miras in $H-K$. The redder colors are due
to photospheric effects and/or circumstellar dust
(Feast et al. 1982; Whitelock et al. 1986; Gaylard et al. 19
89; Glass et al. 1995; Zijlstra et al. 1996)
Note that estimated extinction was determined independent of
the Sgr I Miras in Figure~\ref{cmd}.
We have included stars in Figure~\ref{cmd} from the
list of Glass et al. (1995) with $M_K$ $\leq$ $-$7.0. For clarity,
we have plotted one third of the stars in the actual sample.
Glass et al. derive bolometric magnitudes for these stars
from which we find BC$_K$ $=$ 3.24 $\pm$ 0.15 mag for all the stars in
their sample with $M_K$ $\leq$ $-$7.0.

{\it Baade's window M giants.} The BW M giants are taken from the
list in FW87. This includes variable stars as indicated by FW87 and
the LPVs of Lloyd Evans (1976) which have infrared photometry from
FW87 and Glass et al. (1982). Like the Sgr I LPVs, the BW stars have
estimated $A_K$ which is small ($\sim$ 0.14 mag) and determined
independently of the M giants themselves.

{\it LMC optical LPVs.} This is an analogous data set to the Sgr I
field, but for the LMC.
We have adopted a distance of 46.8 kpc (Reid and Strugnell 1986).
The colors and magnitudes are taken from the large
amplitude variable list of Hughes \& Wood (1990).
We plot in Figure~\ref{cmd} M stars with $M_K$ \aple $-$6.9 mag; for clarity,
not all the stars are plotted. These stars are identified as
Miras by Hughes \& Wood and were discovered optically.
Note that this subset
of LMC Miras reaches higher
luminosities than those in the Galactic bulge (Sgr I field). This is
presumably due to their higher mass (and hence younger age).
We also expect slightly more luminous $M_{\rm bol}$ for the
lower metallicity LMC stars (Vassiliadis \& Wood 1993), but this
difference may be compensated for by the different BC$_K$ for the
two populations.
Using the bolometric and $K$ magnitudes tabulated in Hughes \& Wood
(1990), we find BC$_K$ $=$ 2.94 $\pm$ 0.12 for stars with M spectral
types and $M_K$ $\leq$ 6.9 mag.

{\it LMC IRAS LPVs and supergiants.} These are IRAS selected sources in
the LMC
taken from Wood et al. (1992) and Zijlstra et al. (1996);
they generally have no optical counterparts. Zijlstra et al.
identify candidate AGB stars and supergiants. The LPV stars might be
similar to
optically discovered Miras but with thicker circumstellar dust shells.
Zijlstra et al. classified IRAS sources as supergiants based primarily
on a luminosity criterion, but a small number were classified
based on color and/or small amplitude variability at $K$.
We plot the IRAS sources here mainly to emphasize the
possible range of colors for LPVs. The very red colors for this set are
due to circumstellar dust shells.
The red color is due to a combination of
local extinction in the
shell, excess emission from the dust in the shell, and
photospheric effects in the underlying star.

{\it SMC LA LPVs.} EFH identified a small group of luminous,
large amplitude (LA) variables
in the SMC among their larger sample of supergiants.
Based on the photometric variability of these stars and their very
large H$_2$O absorption (as determined by narrow--band photometry)
EFH classified these variables as lower mass AGB stars, rather than
supergiants (see also Frogel 1983 and Wood et al. 1983).

\section{Appendix 2}
Here we include a brief discussion of the infrared counterparts
to the known OH and \h2o \ masers within our GC field.

Levine et al. (1995) reported detection of an \h2o \ maser within our
field which they
associated with the bright infrared source which we define as IRS 24
(Paper I).
Levine et al. interpreted IRS 24 as a late type M
supergiant based on a qualitative
analysis of its spectrum, \h2o \ maser emission, and derived
$M_K$ from the analysis of
$H$, $K$, and narrow--band $L$ magnitudes. Their observed $H$ and $K$
magnitudes
are within 0.04 mag of the values we reported in Paper I. However,
their adopted extinction law (Rieke et al. 1989) results in a
de--reddened $K$ magnitude which is $\sim$ 0.9 mag brighter than the
value we derive (Table 2). Most of the difference comes from the adopted
$A_L$/$A_K$ in the Rieke et al. extinction law which is larger than the
value in the Mathis (1990) law which we have adopted.

\begin{figure}
\plotone{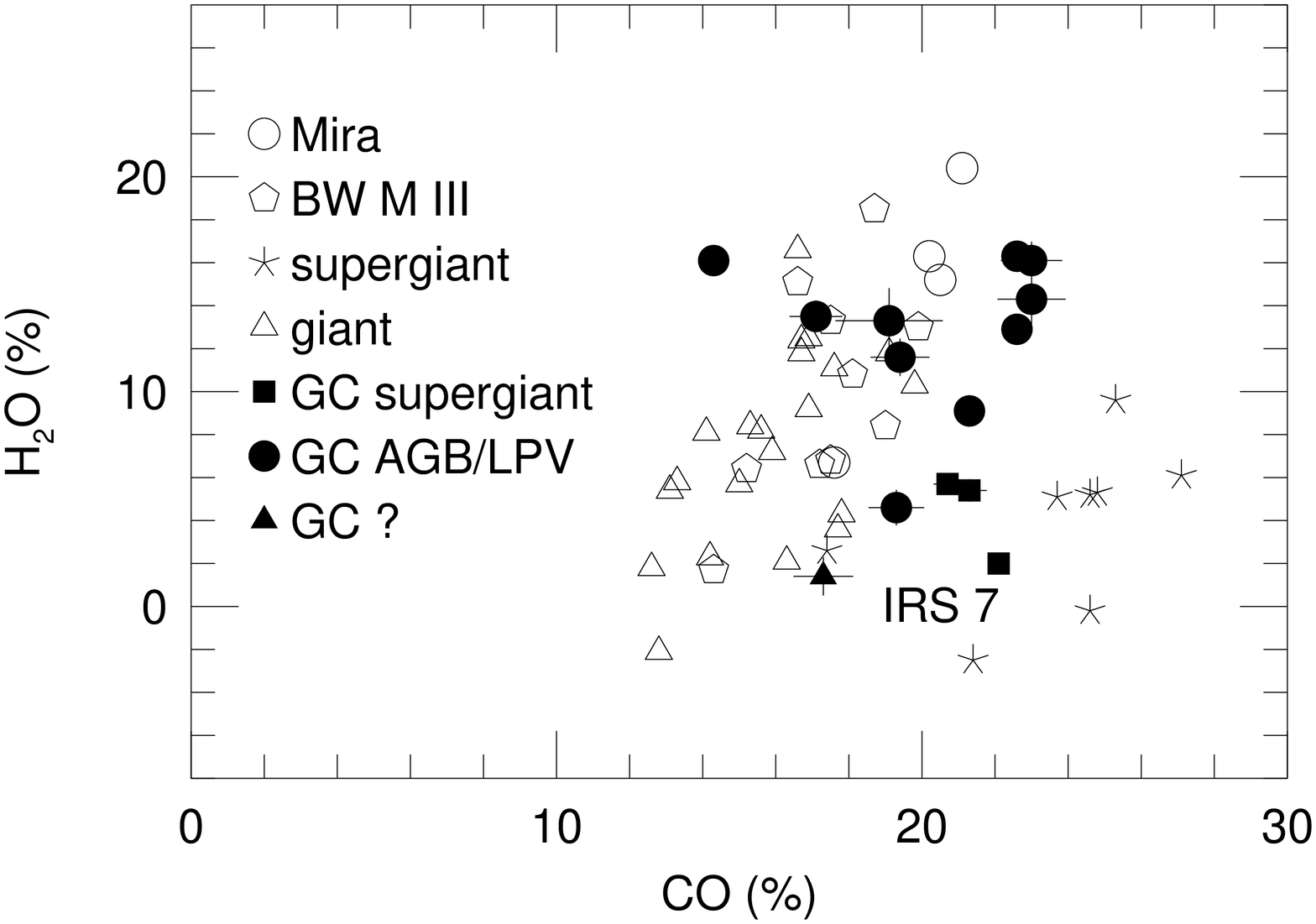}
\caption[]{
Measured  H$_2$O vs. CO strength for Galactic
center (GC) stars ({\it filled symbols})
disk giants ({\it open triangles}),
disk supergiants ({\it asterisks}), Baade's window (BW) M giants
({\it open pentagons}), and disk Miras ({\it open circles}).
The GC stars are identified as supergiants ({\it filled squares}),
asymptotic giant branch (AGB) stars or
long period variables (LPVs) ({\it filled circles}), and
ambiguous ({\it filled triangles}).
The GC data are from the present work,
Sellgren et al. (1987) and Levine et al. (1995).
Only GC stars with $A_K$ determined from two infrared colors
are plotted.
The disk stars are from the present work, Terndrup et al. (1991)
and Kleinmann \& Hall (1986). The Mira variables (or long period
variables, LPVs) are from Johnson \& M\'{e}ndez (1970).
The BW M giants are from Terndrup et al. (1991).
The error bars reflect measurement uncertainty (Table 1) and
uncertainty in $A_K$ (uncertainties listed in Table 2).
The large value of H$_2$O for some of the GC stars suggests they are
LPVs; see text.
}
\label{h2o}
\end{figure}

\begin{figure}
\plotone{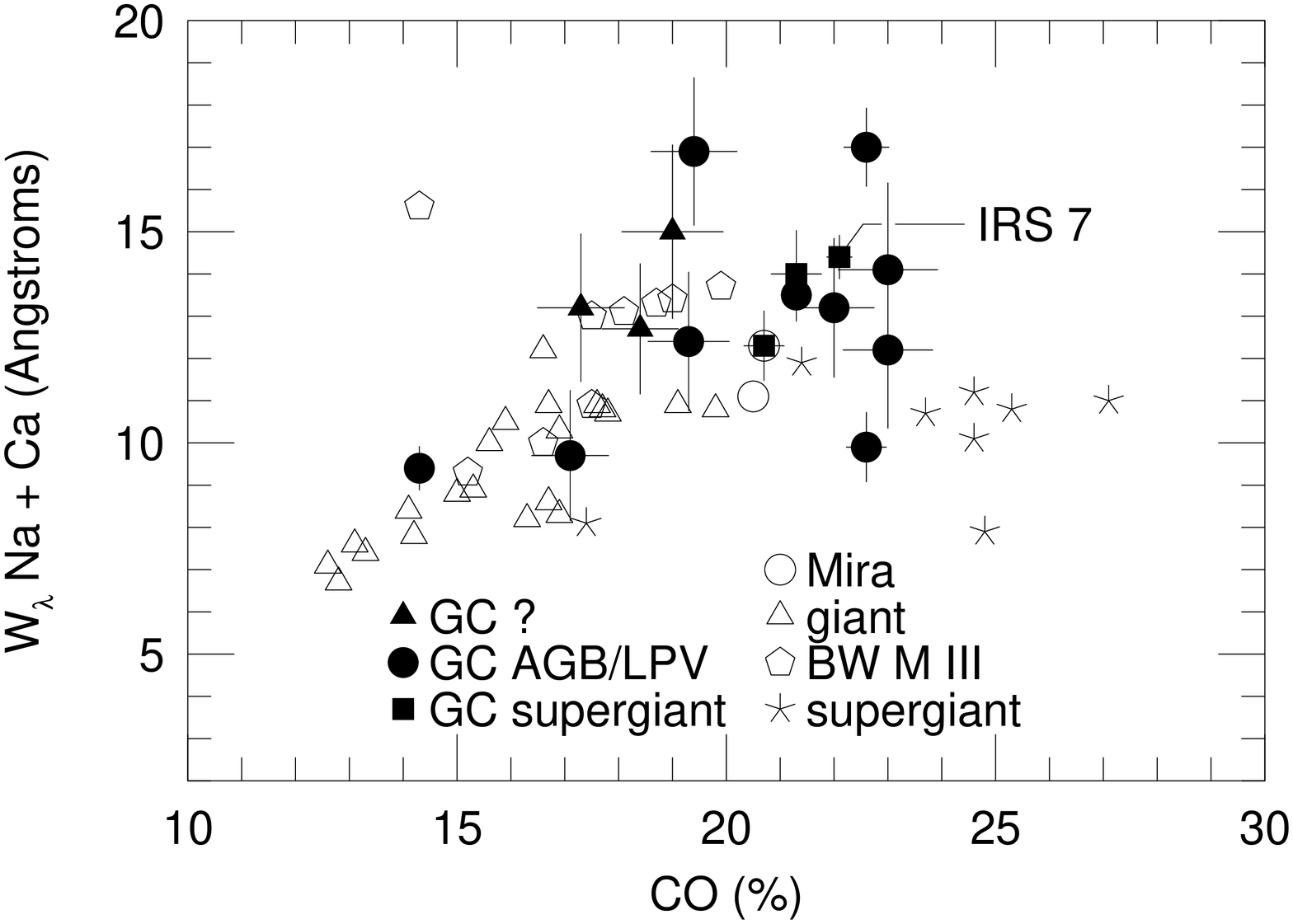}
\caption[]{
Sum of measured ``Na'' and ``Ca'' equivalent
widths vs. CO strength.
Symbols are the same as
for Figure~\ref{h2o}. Competing effects of abundance,
luminosity, and effective temperature all play a role in this diagram.
Galactic center (GC) stars identified as asymptotic giant
branch (AGB) or long period
variables (LPVs) have similar ``Na'' plus ``Ca''
absorption as bulge giants and are enhanced relative to disk giants.
GC stars identified as supergiants (e.g. IRS 7) have significantly
higher ``Na'' plus ``Ca'' than disk supergiants.
}
\label{naca}
\end{figure}

\begin{figure}
\plotfiddle{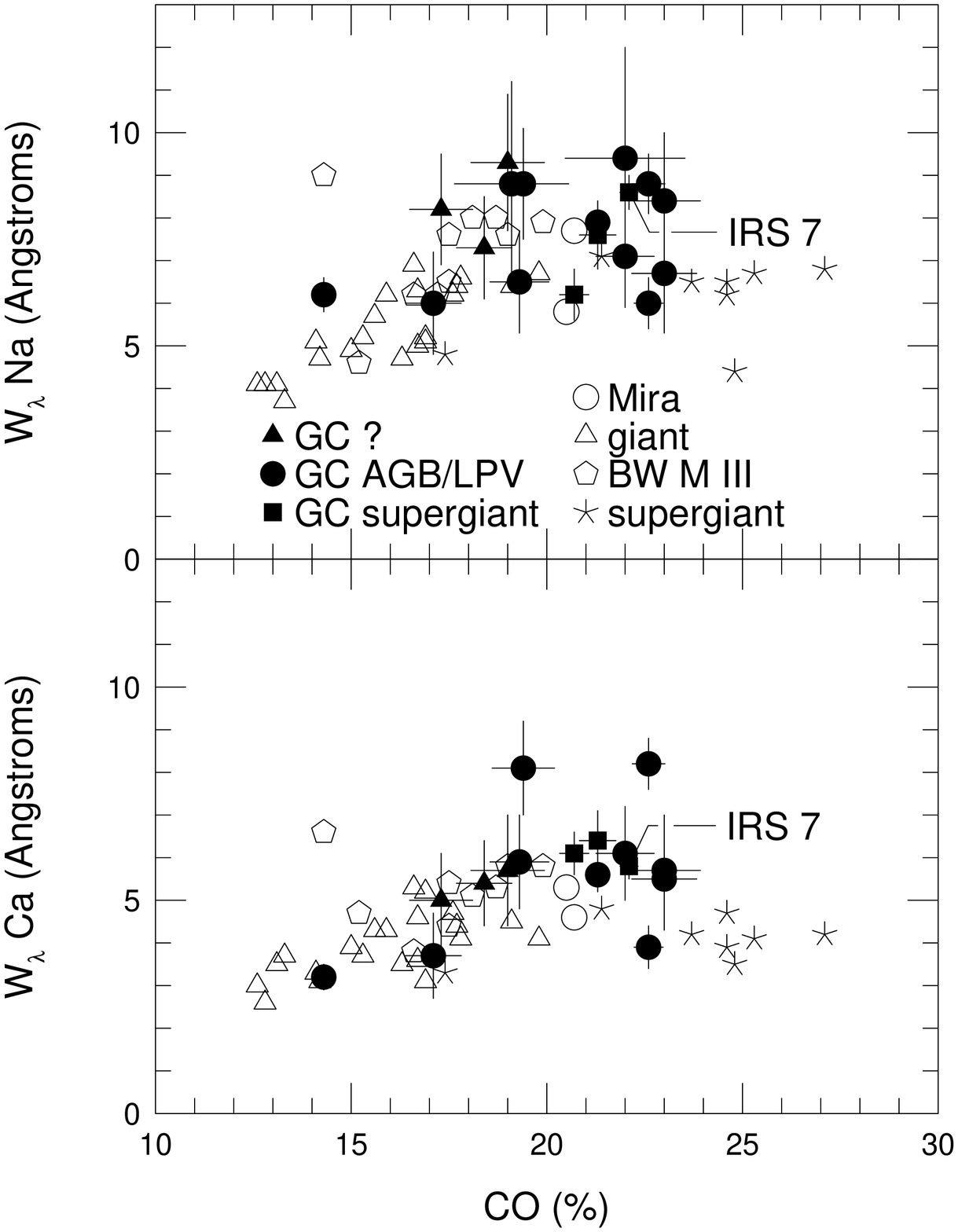}{6.5 in}{0}{70}{70}{-218}{-20}
\caption[]{
``Na'' and ``Ca'' equivalent widths vs. CO strength.
Symbols are the same as for Figures~\ref{h2o} and \ref{naca}.
}
\label{naca2}
\end{figure}

D. Levine and D.
Figer have kindly given us a copy of their IRS 24 spectrum which we have
analyzed along with our other GC spectra (Figure~\ref{spect}, Table 1).
This source has
strong CO and \h2o \ absorption in its $K-$band spectrum (Levine et al.
1995; Table 1). We find that
the derived CO and \h2o \ strengths (Table 2) for IRS 24 are
more consistent with an LPV
classification than a late type supergiant (Figure~\ref{h2o}).
Sjouwerman \& van Langevelde (1996) recently detected 1612 MHz OH emission
associated with IRS 24 and concluded that it is most likely an
intermediate mass LPV star (i.e., it is Mira--like). We believe the
infrared colors, $K$ magnitude and variability, derived $A_K$,
the infrared spectrum, and \h2o \ and OH maser emission are most consistent
with the LPV classification. Although, note that the mass and age we derive
for IRS 24 (Table 2) correspond to a somewhat younger, and more luminous star
than suggested by Sjouwerman \& van Langevelde (1996).

There are four other 1612 MHz OH/IR stars within
our field from the lists of Winnberg et al. (1985) and
Lindqvist et al. (1992). Three of the four
OH/IR stars in our field have infrared counterparts with
$K$ $>$ 10.5 mag. We have no spectra or
variability information for them.

The fourth OH/IR star is of particular interest.
This is star OH359.95--0.05 (Winnberg et al. 1985) or
OH359.946--0.047 (Lindqvist et al. 1992).
This source is within 1$''$ of source IRS 10* which was
identified as a photometric
variable by Tamura et al. (1996). IRS 10* was identified by Blum et al.
(1996, Paper I)
as a very red $K-$band source ($K =$ 10.75 mag) which was not detected
at $H$ or $J$ but which was detected at $L$. Blum et al. (1996) called this
source IRS 10EL. The position of IRS 10* is between IRS 10E and 10W
(Tamura et al. 1996; Blum et al. 1996); the OH/IR star is also positionally
coincident with IRS 10E within the uncertainties ($\sim$ 1$''$ using the
position for
IRS 7 from Tollestrup et al. 1989, the position for OH359.946--0.047
from Lindqvist et al. 1992, the associated uncertainties for both, and the
offsets in Paper I). We believe the very red $K-L$ color of IRS 10*,
as reported in Paper I, strongly enhances the suggestion by
Tamura et al. (1996) that it is the near-infrared counter--part to
OH359.946--0.047 and therefore,  a luminous LPV.

\newpage


\begin{figure}
\caption[]{
Spectra of cool Galactic center (GC) stars,
obtained using OSIRIS with $\lambda / \Delta\lambda = 570$.
All spectra have been corrected for interstellar reddening and
are presented as a normalized ratio of the GC star to an A or B star.
Note the strong absorption longward of 2.3 \mic \ due to CO in all the
stars and strong absorption due to H$_2$O shortward of 2.2 $\mu$m
in some of the stars.
Emission lines are likely due to incomplete subtraction of the
diffuse nebular emission in the GC for some stars
(e.g. IRS 2) but may be associated with other cool stars
(e.g. IRS 9); see text.
The spectra labeled M0 III, M3 III, and M7 III
are examples of our spectral data set
for disk M giants with optically determined spectral types.
IRS 24 is from Levine et al. (1995); R Cas is from Johnson \& M\'endez
(1970). We also show high resolution spectra, after re--binning to
$\lambda / \Delta \lambda$ = 570, of IRS 7 and IRS 23 from Sellgren et
al. (1987) and of $\mu$ Cep (M2 Ia) from Kleinmann \& Hall (1986).
}
\label{spect}
\end{figure}

\begin{figure}
\plotone{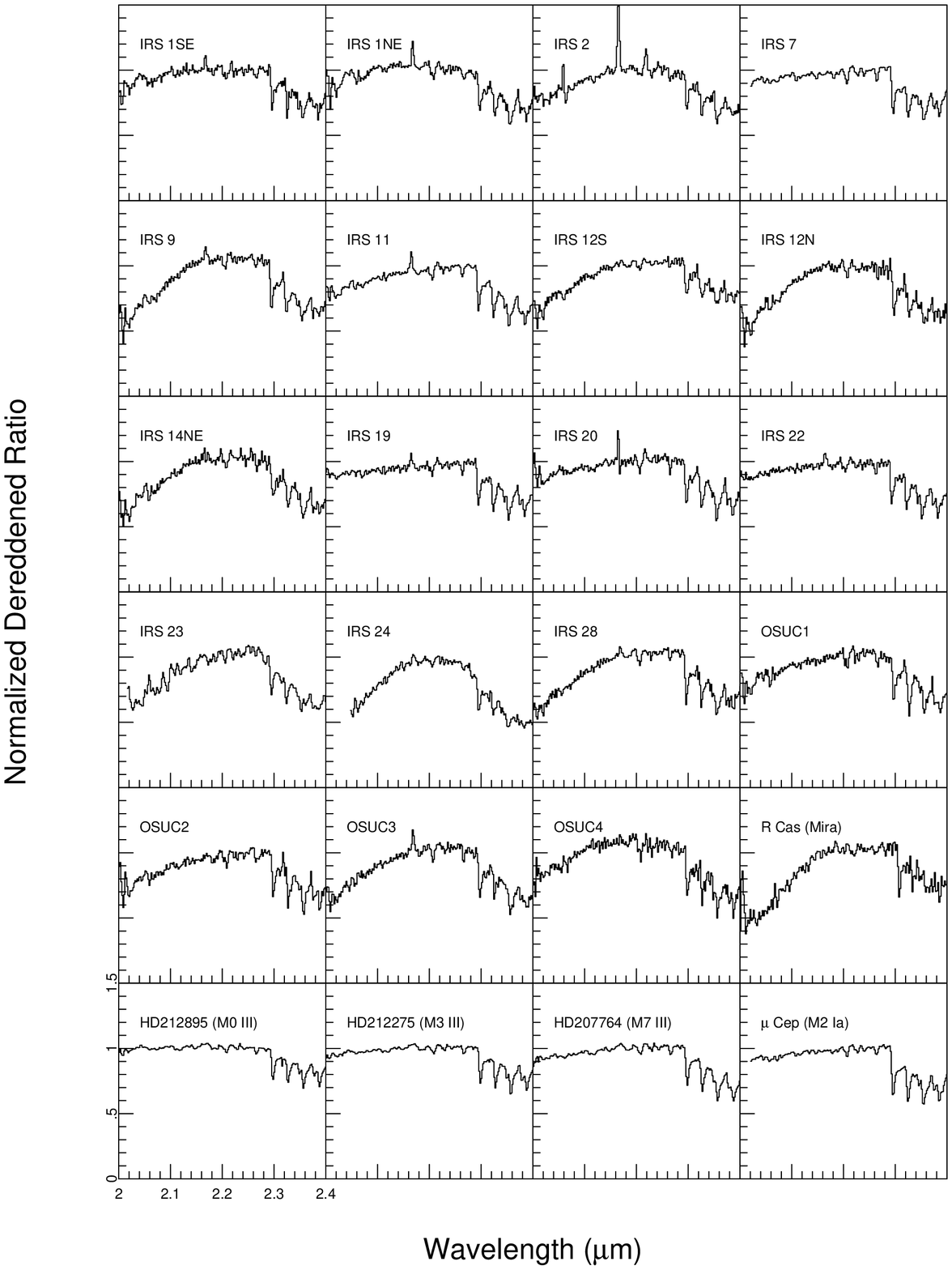}
\end{figure}

\begin{figure}
\caption[]{
Color--magnitude diagram for the Galactic center (GC) cool stars
and comparison stars. Individual data sets are
described in Appendix 1. The GC stars are plotted in each
panel along with known supergiants ({\it left panel}) and
long period variables (LPVs) and M giants
({\it right panel}). IRS 7 is the most luminous GC star in the Figure.
For clarity, not all the fainter stars in the
Baade's window (BW) and LPV data sets are plotted.
A sun-to-Galactic center distance of 8 kpc
(Reid 1993) was adopted for the BW, Sgr I, and GC
stars. The Large Magellanic Cloud (LMC) distance was
taken as 46.8 kpc (Reid \& Strugnell 1986). The error bars reflect
photometric uncertainties in the derived de--reddened $K$ magnitudes
of Blum et al. (1996, Paper I). The effect of one magnitude of
extinction at $K$ is shown by the solid line according to the
interstellar reddening curve of Mathis (1990). The red colors of the
IRAS sources are due to circumstellar dust shells; see Appendix 1.
}
\label{cmd}
\end{figure}

\begin{figure}
\plotone{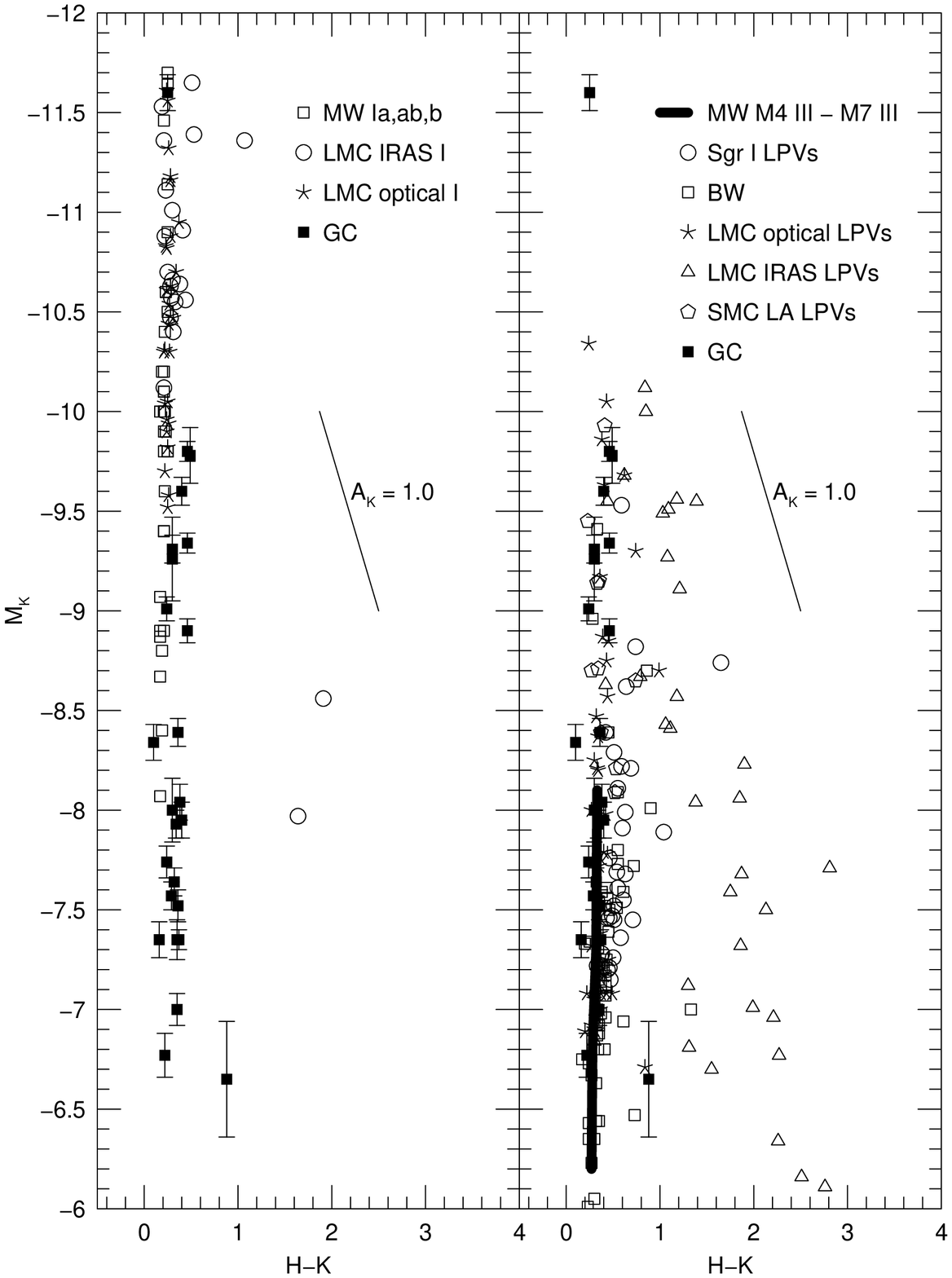}
\end{figure}

\end{document}


\begin{planotable}{lrrrrr}
\tablecaption{Cool Star Absorption Strengths}
\tablenum{1}
\tablehead{
\colhead{Name} &
\colhead{SpType\tablenotemark{a}} &
\colhead{``Na''\tablenotemark{b} \ (\AA)} &
\colhead{``Ca''\tablenotemark{c} \ (\AA)} &
\colhead{CO\tablenotemark{d} \  ($\%$) } &
\colhead{H$_2$O\tablenotemark{e} \  ($\%$) }
}
\startdata
SU Per\tablenotemark{f}       &M3--4 Iab &   6.8 $\pm$   0.2 &   4.2 $\pm$
0.2 &  27.1 $\pm$   0.1 &   6.1 $\pm$   0.2 \nl
KY Cyg\tablenotemark{g}&M3 Ia &   6.7 $\pm$   0.3 &   4.1 $\pm$   0.3 &  25.3 
$\pm$   0.2 & 9.6 $\pm$   0.2\nl
SAO119\tablenotemark{g}&M3 Ia &   7.1 $\pm$   0.3 &   4.8 $\pm$   0.3 &  21.4 
$\pm$   0.2 & $-$2.5 $\pm$   0.2\nl
PZ Cas\tablenotemark{g}&M4 Ia &   6.5 $\pm$   0.4 &   4.2 $\pm$   0.3 &  23.7 
$\pm$   0.2 & 5.1 $\pm$   0.3\nl
RT Car&M2+ Ia&   4.4 $\pm$   0.8 &   3.5 $\pm$   0.7 &  24.8 $\pm$   0.5 &
5.3 $\pm$   0.5\nl
$\alpha$ Ori\tablenotemark{h,i}   &M2 Iab& 6.5 $\pm$ 0.0 &4.7 $\pm$ 0.0& 24.6
$\pm$ 1.2 & $-$0.2 $\pm$ 1.3\nl
$\mu$ Cep\tablenotemark{f}    &M2 Ia &   6.2 $\pm$   0.2 &   3.9 $\pm$   0.2 &
24.6 $\pm$   0.1 &   5.2 $\pm$   0.2 \nl
HR 8726\tablenotemark{f}      &K5 Ib &   4.8 $\pm$   0.2 &   3.3 $\pm$   0.2 &
17.4 $\pm$   0.1 &   2.6 $\pm$   0.2 \nl
R Leo\tablenotemark{h}&M8e, Mira&\nodata&\nodata& 20.2 $\pm$ 0.5 & 16.3 $\pm$
0.5 \nl
R Cas\tablenotemark{h}&M7e, Mira&\nodata&\nodata& 21.1 $\pm$ 0.8 & 20.4 $\pm$
0.8 \nl
R Hyd\tablenotemark{h}&M7e, Mira&\nodata&\nodata& 17.6 $\pm$ 0.8 & 6.7 $\pm$
0.8 \nl
$o$ Cet\tablenotemark{h,i}&M6e, Mira&5.8 $\pm$ 0.0&5.3 $\pm$ 0.0& 20.5 $\pm$
0.8 & 15.2 $\pm$ 0.8 \nl
$\chi$ Cyg\tablenotemark{h,i}&S6$^+$/1e, Mira&7.7 $\pm$ 0.0&4.6 $\pm$ 0.0& 
20.5 $\pm$ 0.8 & \nodata\nl
BK Vir\tablenotemark{f}       &M7$-$ III &   6.7 $\pm$   0.2 &   4.1 $\pm$
0.2 &  19.8 $\pm$   0.1 &  10.3 $\pm$   0.1 \nl
SW Vir\tablenotemark{f}       &M7 III &   6.2 $\pm$   0.2 &   4.7 $\pm$   0.2 &
 17.6 $\pm$   0.1 &  11.1 $\pm$   0.1 \nl
HD 207764    &M7 III &   6.6 $\pm$   0.5 &   4.1 $\pm$   0.4 &  17.8 $\pm$
0.3 &   4.3 $\pm$   0.3 \nl
R Lyra\tablenotemark{f}       &M5 III &   6.4 $\pm$   0.2 &   4.4 $\pm$   0.2 &
 17.7 $\pm$   0.1 &   3.6 $\pm$   0.2 \nl
HD 212275    &M3 III &   4.7 $\pm$   0.4 &   3.5 $\pm$   0.3 &  16.3 $\pm$
0.2 &   2.1 $\pm$   0.3 \nl
$\chi$ Peg\tablenotemark{f}     &M2 III &   4.7 $\pm$   0.2 &   3.1 $\pm$   0.2
&  14.2 $\pm$   0.1 &   2.3 $\pm$   0.2 \nl
$\gamma$ Sge\tablenotemark{f}   &M0 III &   4.1 $\pm$   0.2 &   3.0 $\pm$   0.2
&  12.6 $\pm$   0.1 &   1.8 $\pm$   0.2 \nl
HD 212895    &M0 III &   4.1 $\pm$   0.4 &   2.6 $\pm$   0.3 &  12.8 $\pm$
0.2 & $-$2.1 $\pm$   0.3 \nl
HD 89060\tablenotemark{j}     &M4 III &   6.2 $\pm$   0.3 &   4.3 $\pm$   0.3 &
 15.9 $\pm$   0.2 &   7.2 $\pm$   0.2 \nl
HD 89951\tablenotemark{j}     &M3 III &   4.9 $\pm$   0.2 &   3.9 $\pm$   0.2 &
 15.0 $\pm$   0.1 &   5.7 $\pm$   0.2 \nl
HD 94152\tablenotemark{j}     &M6 III &   5.2 $\pm$   0.2 &   3.1 $\pm$   0.2 &
 16.9 $\pm$   0.1 &  12.5 $\pm$   0.1 \nl
HD 99495\tablenotemark{j}     &M4 III &   5.7 $\pm$   0.3 &   4.3 $\pm$   0.3 &
 15.6 $\pm$   0.2 &   8.2 $\pm$   0.2 \nl
HD 100569\tablenotemark{j}    &M2 III &   3.7 $\pm$   0.2 &   3.7 $\pm$   0.2 &
 13.3 $\pm$   0.1 &   5.8 $\pm$   0.2 \nl
HD 100783\tablenotemark{j}    &M2 III &   4.1 $\pm$   0.2 &   3.5 $\pm$   0.2 &
 13.1 $\pm$   0.1 &   5.4 $\pm$   0.2 \nl
HD 102608\tablenotemark{j}    &M7 III &   6.3 $\pm$   0.2 &   4.6 $\pm$   0.2 &
 16.7 $\pm$   0.1 &  12.4 $\pm$   0.1 \nl
HD 102766\tablenotemark{j}    &M6 III &   6.4 $\pm$   0.2 &   4.5 $\pm$   0.2 &
 19.1 $\pm$   0.1 &  11.8 $\pm$   0.1 \nl
HD 104745\tablenotemark{j}    &M3 III &   5.0 $\pm$   0.2 &   3.6 $\pm$   0.2 &
 16.7 $\pm$   0.1 &  11.8 $\pm$   0.1 \nl
HD 109225\tablenotemark{j}    &M5 III &   5.1 $\pm$   0.3 &   3.3 $\pm$   0.3 &
 14.1 $\pm$   0.2 &   8.1 $\pm$   0.2 \nl
HD 109467\tablenotemark{j}    &M6 III &   5.2 $\pm$   0.3 &   3.7 $\pm$   0.3 &
 15.3 $\pm$   0.2 &   8.4 $\pm$   0.2 \nl
RT Vir\tablenotemark{j}       &M8 III &   6.9 $\pm$   0.3 &   5.3 $\pm$   0.3 &
 16.6 $\pm$   0.2 &  16.6 $\pm$   0.2 \nl
HD 126903\tablenotemark{j}    &M6 III &   5.1 $\pm$   0.5 &   5.2 $\pm$   0.5 &
 16.9 $\pm$   0.3 &   9.2 $\pm$   0.3 \nl
BW 4-028\tablenotemark{j}     &M7 III &   6.2 $\pm$   0.3 &   3.8 $\pm$   0.3 &
 16.6 $\pm$   0.2 &  15.1 $\pm$   0.2 \nl
BW 4-055\tablenotemark{j}     &M8 III &   8.0 $\pm$   0.4 &   5.3 $\pm$   0.3 &
 18.7 $\pm$   0.2 &  18.5 $\pm$   0.2 \nl
BW 4-78\tablenotemark{j}      &M5 III &   4.6 $\pm$   0.2 &   4.7 $\pm$   0.2 &
 15.2 $\pm$   0.1 &   6.4 $\pm$   0.2 \nl
BW 4-93\tablenotemark{j}      &M6 III &   6.5 $\pm$   0.5 &   4.4 $\pm$   0.5 &
 17.5 $\pm$   0.3 &   6.8 $\pm$   0.4 \nl
BW 4-107\tablenotemark{j}     &M6 III &   9.0 $\pm$   0.1 &   6.6 $\pm$   0.1 &
 14.3 $\pm$   0.0 &   1.7 $\pm$   0.1 \nl
BW 4-133\tablenotemark{j}     &M6 III &   6.2 $\pm$   0.5 &  \nodata &  17.2
$\pm$   0.3 &   6.6 $\pm$   0.4 \nl
BW 4-165\tablenotemark{j}     &M7 III &   7.6 $\pm$   0.6 &   5.8 $\pm$   0.5 &
 19.0 $\pm$   0.4 &   8.4 $\pm$   0.4 \nl
BW 4-179\tablenotemark{j}     &M7 III &   8.0 $\pm$   0.2 &   5.1 $\pm$   0.2 &
 18.1 $\pm$   0.1 &  10.8 $\pm$   0.1 \nl
BW 4-247\tablenotemark{j}     &M8 III &   7.6 $\pm$   0.3 &   5.4 $\pm$   0.3 &
 17.5 $\pm$   0.2 &  13.3 $\pm$   0.2 \nl
BW 4-289\tablenotemark{j}     &M9 III &   7.9 $\pm$   0.5 &   5.8 $\pm$   0.4 &
 19.9 $\pm$   0.3 &  13.0 $\pm$   0.3 \nl
IRS 1NE     & K5--M0 I or M7 III&   7.3 $\pm$   1.2 &   5.4 $\pm$   1.0 &  18.4 
$\pm$ 0.7 & \nodata \nl
IRS 1SE     & K5 I or M7 III &   8.2 $\pm$   1.3 &   5.0 $\pm$   1.1 &  
17.3 $\pm$ 0.8 &  1.4 $\pm$   0.9 \nl
IRS 2  & K5 I or M7 III &  \nodata &  \nodata &  16.7 $\pm$   0.9 & \nodata \nl
IRS 7\tablenotemark{k}     & M1 I&   8.6 $\pm$   0.4 &   5.8 $\pm$   0.3 &
22.1 $\pm$   0.2 &   2.0 $\pm$   0.3 \nl
IRS 9        & $>$ M7 III &   6.7 $\pm$   1.4 &   5.5 $\pm$   1.2 &  23.0 $\pm$
0.8 &  16.1 $\pm$   0.9 \nl
IRS 11       & $>$ M7 III &   7.9 $\pm$   0.5 &   5.6 $\pm$   0.4 &  21.3 $\pm$
0.3 &   9.1 $\pm$   0.3 \nl
IRS 12N     & $>$ M7 III &   8.4 $\pm$   1.6 &   5.7 $\pm$   1.3 &  23.0 $\pm$
0.9 &  14.3 $\pm$   1.5 \nl
IRS 12S     & M6 III&   6.0 $\pm$   1.2 &   3.7 $\pm$   1.0 &  17.1 $\pm$
0.7 &  13.5 $\pm$   0.7 \nl
IRS 14NE    & M7 III&   8.8 $\pm$   2.4 &  \nodata &  19.1 $\pm$   1.5 &
13.3 $\pm$   1.5 \nl
IRS 19\tablenotemark{k} & M0 I&   6.2 $\pm$   0.6 &   6.1 $\pm$   0.5
&  20.7 $\pm$   0.4 &   5.7 $\pm$   0.4 \nl
IRS 20       & M7 III&   6.5 $\pm$   1.2 &   5.9 $\pm$   1.1 &  19.3 $\pm$
0.8 &   4.6 $\pm$   0.8 \nl
IRS 22\tablenotemark{k} &M1 I &   7.6 $\pm$   0.8 &   6.4 $\pm$   0.7
&  21.3 $\pm$   0.5 &   5.4 $\pm$   0.5 \nl
IRS 23\tablenotemark{k} & $>$ M7 III&   6.2 $\pm$   0.4 &   3.2 $\pm$   0.3 &
14.3 $\pm$   0.2 &  16.1 $\pm$   0.2 \nl
IRS 24\tablenotemark{l} & $>$ M7 III &   6.0 $\pm$   0.6 &   3.9 $\pm$   0.5 &
22.6 $\pm$   0.4 &  12.9 $\pm$ 0.4\nl
IRS 28       & $>$ M7 III&   8.8 $\pm$   0.7 &   8.2 $\pm$   0.6 &  22.6 $\pm$ 
0.4 &  16.3 $\pm$   0.4 \nl
OSU C1 & M7 III&   9.3 $\pm$   1.6 &   5.7 $\pm$   1.3 &  19.0 $\pm$
0.9 & \nodata \nl
OSU C2  & M7 III&   8.8 $\pm$   1.3 &   8.1 $\pm$   1.1 &  19.4 $\pm$ 0.8 
&  11.6 $\pm$   0.9 \nl
OSU C3 & M0--1 I or $>$ M7 III &   7.1 $\pm$   1.2 &   6.1 $\pm$   1.1 &  
22.0 $\pm $ 0.7 &  \nodata\nl
OSU C4       & $>$ M7 III &   9.4 $\pm$   2.6 &  \nodata &  22.0 $\pm$   1.5 &
\nodata \nl
\enddata
\tablenotetext{a}{Spectral type taken from same reference as spectrum for disk
and bulge stars. RT Car spectral type from Humphreys (1978). 
OSIRIS M III spectral
types are from the Michigan Spectral Catalog.
For the Galactic center stars, spectral type is estimated based on the
analysis of the Galactic center
spectra and $M_K$. M III refers to both AGB and LPV
stars; see text, Table 2.}
\tablenotetext{b}{Equivalent width (W$_\lambda$) 
in 0.015 $\mu$m wide band centered at 2.206 \mic,
near the Na I doublet. ``Na'' refers to the fact that this feature is
near the Na doublet, but more than half of the W$_\lambda$ may be due to
Sc, Ti, Si, and V; see text.}
\tablenotetext{c}{W$_\lambda$ in 0.013 $\mu$m wide band centered at 2.264 \mic,
near the Ca I triplet. ``Ca'' refers to the fact that this feature is
near the Ca triplet, but more than half of the W$_\lambda$ may be due to
Fe, Ti, and Sc; see text.}
\tablenotetext{d}{CO 2--0 rotational vibrational band. Absorption strength is
defined as
$(1 - F_{CO}/F_{cont}) \times 100$, where $F_{CO}$ and $F_{cont}$ are the
integrated flux
in 0.015 $\mu$m bands centered at 2.302 $\mu$m and 2.284 \mic, respectively.}
\tablenotetext{e}{H$_2$O absorption strength defined as
$(1 - F_{H_2O}/F_{cont}) \times 100$, where $F_{H_2O}$ and $F_{cont}$ are the
integrated flux
in 0.015 $\mu$m bands located at 2.095 $\mu$m and 2.284 $\mu$m,
respectively. Only Galactic center 
stars with $A_K$ determined from two or more infrared
colors (Paper I) have measured H$_2$O.}
\tablenotetext{f}{Spectrum taken from Kleinmann \& Hall (1986) and re--binned
to 19.4 \AA \
pix$^{-1}$; see text.}
\tablenotetext{g}{Spectrum taken from Hanson et al. (1996). Spectra
are $\sim$ 16 \AA \ pix$^{-1}$, not re--binned.}
\tablenotetext{h}{Spectrum taken from Johnson \& M\'{e}ndez (1970). Spectra
are $\sim$ 40 \AA \ pix$^{-1}$, not re--binned.}
\tablenotetext{i}{``Na'' and ``Ca'' measured from high resolution
spectra of Wallace \& Hinkle (1996).}
\tablenotetext{j}{Spectrum taken from Terndrup et al. (1991). Spectra are 22
\AA \ pix$^{-1}$, not re--binned. 
BW stars have grism spectral types which may be different than
MK types; see Terndrup et al. (1990).}
\tablenotetext{k}{Spectra taken from Sellgren et al. (1987) and re--binned to
19.4 \AA \
pix$^{-1}$; see text. Absorption strengths for IRS 19 and 22 are unweighted
averages
of values derived from the OSIRIS spectra and re--binned Sellgren et al.
spectra.}
\tablenotetext{l}{Spectrum taken from Levine et al. (1995). Spectrum is 16 \AA \
\pix, not re--binned.}
\tablecomments{Spectra are from this paper unless otherwise noted.}
\end{planotable}


\begin{deluxetable}{lrrrrrcccc}
\scriptsize
\tablecaption{Galactic Center Star Properties}
\tablenum{2}
\tablehead{
\colhead{Name} &
\colhead{$M_K$\tablenotemark{a}} &
\colhead{$(H-K)_\circ$\tablenotemark{b}} &
\colhead{$A_K$\tablenotemark{c}} &
\colhead{$M_{Kfaint}$\tablenotemark{d}} &
\colhead{Type\tablenotemark{e}} &
\colhead{$M_{\rm bol}$\tablenotemark{f}} &
\colhead{\teff\tablenotemark{g} \ (K)} &
\colhead{M (M$_\odot$)\tablenotemark{h}} &
\colhead{Age (Myr)\tablenotemark{h}}
}
\startdata
IRS 7  &$-$11.60 $\pm$ 0.09 & 0.23 $\pm$ 0.11 & 3.48 $\pm$ 0.09&$-$10.12&
SG&$-$9.0 &3600& 20$-$25 & 7$-$9\nl
IRS 19 &$-$9.80 $\pm$ 0.05& 0.46 $\pm$ 0.06& 3.50 $\pm$ 0.04&$-$8.30& 
SG&$-$7.2 &3660& 12$-$15 & 12$-$18 \nl
IRS 12N&$-$9.78 $\pm$ 0.14& 0.49 $\pm$ 0.11 & 3.79  
$\pm$ 0.14&$-$7.99& LPV& $-$6.5&2600 & $\sim$ 5 & $\sim$ 100 \nl
IRS 24\tablenotemark{i} &$-$9.60 $\pm$ 0.07& 0.40 $\pm$ 0.08 & 3.34 $\pm$
0.06&$-$8.26& LPV& $-$6.4 &2600&$\sim$ 5 &$\sim$  100 \nl
IRS 23 &$-$9.34 $\pm$ 0.05& 0.46 $\pm$ 0.06& 3.44 $\pm$ 0.04&$-$7.90& LPV
&$-$6.1 &2600& $\sim$ 4 & $\sim$ 200 \nl
IRS 9  &$-$9.31 $\pm$ 0.07& 0.30 $\pm$ 0.07 & 3.36 $\pm$ 0.06&$-$7.95& LPV&
$-$6.1 &2600& $\sim$ 4 & $\sim$ 200 \nl
IRS 2  &$-$9.26 $\pm$ 0.21& 0.30 $\pm$ 0.18 & 5.31 $\pm$ 0.20&$-$5.95& ?& 
$-6.8 $ -- $ -$6.1 & 3750 -- 3210 &?& ? \nl
IRS 22 &$-$9.01 $\pm$ 0.06& 0.24 $\pm$ 0.07 & 2.52 $\pm$ 0.05&$-$8.49&
SG&$-$6.4 &3600& 9$-$12 & 18$-$29\nl
IRS 28 &$-$8.90 $\pm$ 0.06& 0.46 $\pm$ 0.06 &  3.74 $\pm$ 0.05&$-$7.16&
LPV&$-$5.7 &2800& $\sim$4 & $\sim$200\nl
IRS 14NE&$-$8.39 $\pm$ 0.07& 0.36 $\pm$ 0.07 &3.61 $\pm$ 0.05&$-$6.78&
AGB&$-$5.2 &3210& 2 -- 4 & 190 -- 1600 \nl
IRS 11&$-$8.34 $\pm$ 0.09 & 0.10 $\pm$ 0.09& 3.00 $\pm$ 0.06&$-$7.34& AGB
&$-$5.1 &$<$ 3210& 2 -- 4 & 190 -- 1600 \nl
OSU C3&$-$8.04 $\pm$ 0.09& 0.38 $\pm$ 0.08&  4.25 $\pm$ 0.09&$-$5.79& ?&
$-5.5$ -- $-$4.8 &  3600 -- 3210&?& ? \nl
IRS 1NE&$-$8.00 $\pm$ 0.16& 0.30 $\pm$ 0.13 & 3.48 $\pm$ 0.14&$-$6.52& ?& 
$-5.5$ -- $-$4.8& 3750 -- 3210& ?&? \nl
F95 B  &$-$7.95 $\pm$ 0.09& 0.40 $\pm$ 0.09 & 3.32 $\pm$ 0.07&$-$6.63& \nodata&
\nodata&\nodata&\nodata&\nodata\nl
IRS 14SW&$-$7.93 $\pm$ 0.07& 0.34 $\pm$ 0.07 &3.56 $\pm$ 0.06&$-$6.37& \nodata&
\nodata&\nodata&\nodata&\nodata\nl
IRS 1SE&$-$7.74 $\pm$ 0.08 & 0.24 $\pm$ 0.07& 3.47 $\pm$ 0.07&$-$6.27&
?& $-5.2$ -- $-4.5$&3750 -- 3210& ? &  ? \nl 
OSU C2&$-$7.64 $\pm$ 0.07 & 0.32 $\pm$ 0.07 & 3.22 $\pm$ 0.06&$-$6.42&
AGB&$-$4.4 & 3210& 1 -- 3& 440 -- 12000 \nl
IRS 12S&$-$7.57 $\pm$ 0.07 & 0.29 $\pm$ 0.07& 3.00 $\pm$ 0.06&$-$6.57&
AGB&$-$4.3 &3380& 1 -- 3& 440 -- 12000 \nl
OSU C4&$-$7.52 $\pm$ 0.08 & 0.36 $\pm$ 0.06 & 3.67 $\pm$ 0.07&$-$5.85&
AGB&$-$4.3 &$<$ 3210& 1 -- 3& 440 -- 12000 \nl
OSU C1&$-$7.35 $\pm$ 0.10& 0.35 $\pm$ 0.09& 3.46 $\pm$ 0.10 &$-$5.89&
AGB&$-$4.1 & 3210& 1 -- 3& 440 -- 12000 \nl
IRS 10E&$-$7.35 $\pm$ 0.09& 0.16 $\pm$ 0.09 & 3.19 $\pm$ 0.07&$-$6.16&
\nodata&\nodata&\nodata&\nodata&\nodata\nl
F95 J  &$-$7.35 $\pm$ 0.05& 0.37 $\pm$ 0.05 & 3.21 $\pm$ 0.04&$-$6.14&
\nodata&\nodata&\nodata&\nodata&\nodata\nl
IRS 20&$-$7.00 $\pm$ 0.08& 0.35 $\pm$ 0.08 & 3.06 $\pm$ 0.06 &$-$5.94&
AGB&$-$3.8 &3210&$<$ 3& $>$440 \nl
IRS 33W&$-$6.77 $\pm$ 0.11& 0.22 $\pm$ 0.11 & 3.12 $\pm$ 0.09 &$-$5.65&
\nodata&\nodata&\nodata&\nodata&\nodata\nl
IRS 29S&$-$6.65 $\pm$ 0.29 & 0.88 $\pm$ 0.29& 2.68  $\pm$ 0.08&$-$5.97&
\nodata&\nodata&\nodata&\nodata&\nodata\nl
\enddata
\tablenotetext{a}{De--reddened $K$ magnitude taken from Blum et al.
(1996, Paper I). Uncertainty is a combination of photometric uncertainty
in observed $K$ and uncertainty in $A_K$ (itself due to photometric
uncertainty) added in quadrature. Distance to Galactic center taken
as 8 kpc (Reid 1993).}
\tablenotetext{b}{Taken from Blum et al. (1996, Paper I).
$(H-K)_\circ$ follows from the observed $H-K$, the adopted extinction law
(which gives the $E(H-K$) in terms of $A_K$), and
the derived $A_K$. $A_K$ was first determined by assuming typical
intrinsic $J-H$ and $H-K$ for M stars (IRS 1NE, 2, OSU C1, C3, and C4 had
only $H-K$ measured).
IRS 29S may have an excess at $H-K$
relative to $J-H$; see Blum et al. (1996, Paper I).}
\tablenotetext{c}{Derived in Blum et al. (1996, Paper I) by assuming
intrinsic colors for M stars; see note b. Uncertainty is photometric
uncertainty only.}
\tablenotetext{d}{Lower luminosity limit based on observed $K$ and
minimum $A_K$; see Appendix 1.}
\tablenotetext{e}{Classification: SG, supergiant; LPV, shows strong
\h2o \ and photometric variability; AGB, based on $M_K$, CO, and \h2o; ?,
$M_K$ and CO are consistent with either a SG or AGB classification.
Spectral types are given in Table 1.
Note that four stars have no spectrum analyzed in this paper. These
four are identified as cool stars by Krabbe et al. (1995) and Figer (1995).
Stars with ``?'' classifications are discussed further in \S~4.}
\tablenotetext{f}{Bolometric magnitude derived using BC$_K$ $=$ 2.6 and 3.2
for supergiants and LPV/AGB stars, respectively; see text and Appendix 1.
Stars with ``?'' classifications have a range of bolometric magnitude
corresponding to supergiant or AGB classification.}
\tablenotetext{g}{Effective temperature, \teff,
from estimated spectral type (Table~1). See text
for adopted \teff \ vs. spectral type relations. Galactic center 
LPVs have \teff \
estimated from $M_{\rm bol}$ and mass; see text.
Stars with ``?'' classifications have a range of effective temperature
corresponding to supergiant or AGB classification.}
\tablenotetext{h}{Mass and age estimated from evolutionary models; see text.
Values of mass and age for stars with ``?'' classifications are disscussed
in the text.}
\tablenotetext{i}{IRS 24 has been identified as an \h2o \ and OH maser
source; see Appendix 2.}
\end{deluxetable}